\documentclass[12pt]{article}
\usepackage{amsmath,amssymb}
\usepackage{bm}
\usepackage{color}
\usepackage{cite}
\usepackage{mathtools}

\usepackage{ulem}
\usepackage{xcolor}
\newcommand{\teal}[1]{{\color{teal}{#1}}}

\usepackage{multirow}

\newcommand{\CL}[2]{\mathcal{C}_{\underset{#2}{#1}}}

\newcommand{\brackets}[1]{\left( #1 \right)}

\def\Slash#1{\rlap/#1}
\def\Slash#1{\rlap{\kern .1em /}#1}

\usepackage{here}
\usepackage {cancel}

\begin{document}

	\title{
	\begin{flushright}
		\ \\*[-80pt]
		\begin{minipage}{0.2\linewidth}
			\normalsize
		\end{minipage}
	\end{flushright}
	%
	{\Large \bf
		Leptonic dipole operator with $\Gamma_2$ modular
		 invariance  
		\\*[10pt]
	in light of Muon $(g-2)_\mu$
		\\*[20pt]}
}

\author{
  Takaaki Nomura $^{1}$\footnote{email: nomura@scu.edu.cn},	Morimitsu Tanimoto $^{2}$\footnote{email: tanimoto@muse.sc.niigata-u.ac.jp}, Xing-Yu Wang $^{1}$\footnote{email: xingyuwang@stu.scu.edu.cn}
	\hskip 0.5 cm 
	\\*[20pt]
	\centerline{
		\begin{minipage}{\linewidth}
			\begin{center}
					$^1${\it \normalsize
					College of Physics, Sichuan University, Chengdu 610065, China} \\
					$^2${\it \normalsize
					Department of Physics, Niigata University, Niigata 950-2181, Japan }
				\\*[10pt]
			\end{center}
	\end{minipage}}
	\\*[20pt]}
\date{\today\\	\vskip 0.8 cm
{\small \bf Abstract}
	\begin{minipage}{0.9\linewidth}
		\medskip \medskip
		\small
We have studied the leptonic EDM and the LFV decays relating with the recent data of anomalous magnetic moment of muon, $(g-2)_{\mu}$ in the leptonic dipole operator. We have adopted the successful $\Gamma_2$ modular invariant model by Meloni-Parriciatu as the flavor symmetry of leptons. Suppose the anomaly of $(g-2)_{\mu}$, $\Delta a_{\mu}$ to be evidence of New Physics (NP), we have related it with the anomalous magnetic moment of the electron $\Delta a_e$, the electron EDM $d_e$ and the $\mu\to e \gamma$ decay. We found that the NP contributions to $\Delta a_{e(\mu)}$ are proportional to the lepton masses squared likewise the naive scaling $\Delta a_\ell \propto m^2_\ell$. The experimental constraint of $|d_e|$ is much tight compared with the one from the branching ratio $\mathcal{B} (\mu \to e \gamma)$ in our framework. Supposing the phase of our model parameter $\delta_{\alpha}$ for the electron to be of order one, we have estimated the upper-bound of $\mathcal{B}(\mu \to e \gamma)$, which is at most $10^{-21}-10^{-20}$. If some model parameters are real, leptonic EDMs vanish since the CP phase of the modular form due to modulus $\tau$ does not contribute to the EDM. However, we can obtain $\mathcal{B} (\mu \to e \gamma)\simeq 10^{-13}$ with non-vanishing $d_e$ in a specific case. The imaginary part of a parameter can lead to $d_e$ in the next-to-leading contribution. The predicted electron EDM is below $10^{-32}$e\,cm, while $\mathcal{B} (\mu \to e \gamma)$ is close to the experimental upper-bound.  The branching ratios of $\tau\to e\gamma$ and $\tau\to \mu\gamma$ are also discussed.
	\end{minipage}}

\begin{titlepage}
\maketitle
\thispagestyle{empty}
\end{titlepage}



\newpage

\section{Introduction}
The electric and  magnetic dipole moments  of 
the charged leptons are  low-energy probes 
of New Physics (NP) beyond the Standard Model (SM). 
The muon $(g-2)_\mu$ experiment at Fermilab reported  a new measurement of the muon magnetic anomaly
\cite{Muong-2:2023cdq}.  Improvements of the analysis and run condition  lead to more than a factor of two reduction in the systematic
uncertainties, which is  compared  with
the E989 experiment at Fermilab~\cite{Muong-2:2021ojo} and the previous BNL result~\cite{Muong-2:2006rrc}. 
This result indicates the discrepancy of  $5.1\,\sigma$ with the SM prediction 
\cite{Aoyama:2020ynm}
(see also \cite{Jegerlehner:2017gek,Colangelo:2018mtw,Hoferichter:2019mqg,Davier:2019can,Keshavarzi:2019abf,Hoid:2020xjs,Czarnecki:2002nt, Melnikov:2003xd,Aoyama:2012wk, Gnendiger:2013pva}).

While there is a debatable point 
on the precise value of the SM prediction, 
that is  the contribution of the hadronic vacuum polarization (HVP).
The current situation is still complicated.
The CMD-3 collaboration \cite{CMD-3:2023alj}  released results on the 
cross section that disagree at the $(2.5-5) \sigma$ level with all previous measurements.
The origin of this discrepancy is currently unknown.
The Bell-{I\hskip -0.4mm}I experiment is expected to measure $e^+e^- \to \pi^+\pi^-$ cross section in the near future
\footnote{Belle-{I\hskip -0.4mm}I reported the measurement of the $e^+e^- \to \pi^+\pi^-\pi^0$ cross section in the energy range $0.62-3.50$\,GeV.
	The result differs by $2.5$ standard deviations from the most precise current determination \cite{Belle-II:2024msd}.
}.
The BMW collaboration published the first complete lattice-QCD result with subpercent precision  \cite{Borsanyi:2020mff}.
Their result is closer to the experimental average ($1.7\sigma$ tension
for no NP in the muon $(g-2)_\mu$).
Further studies are underway to clarify these theoretical differences. 


If the muon $(g-2)_\mu$ anomaly comes from NP, it possibly appears
in other observables of the charged lepton sector. 
The interesting one is the electric dipole moments (EDM)  of the electron. 
 A new upper-bound on the electron EDM, which is $|d_e|<4.1 \times 10^{-30} \rm e\, cm \,(90\%\, { confidence})$,
 was reported by the JILA group 
 \cite{Roussy:2022cmp}.
It overcame the latest ACME collaboration result obtained in 2018
\cite{Andreev:2018ayy}.
Precise measurements of  the electron EDM will be rapidly being updated. The future sensitivity
at ACME $\rm I\hskip -0.04cm I\hskip -0.04cm I$ is expected to be
$|d_e|<0.3\times10^{-30}\,\text{e\,cm}$
\cite{Kara:2012ay,ACMEIII}.
On the other hand, the present upper-bound of the muon EDM \cite{Muong-2:2008ebm} and tauon EDM \cite{Belle:2002nla,Bernreuther:2021elu,Uno:2022xau} are not so tight.	

The lepton flavor violation (LFV) is also  possible NP phenomena of 
charged leptons.
The tightest constraint for LFV is the branching ratio
of the $\mu\to e\gamma$ decay.
The recent experimental upper-bound is {$\mathcal{B}({\mu^+ \to e^+ \gamma}) < 3.1 \times 10^{-13}$ by the combination of the MEG and MEG II experiments~\cite{TheMEG:2016wtm,MEGII:2023ltw}.}
In contrast, the present upper-bounds of
$\mathcal{B}\,({\tau \to \mu \gamma})$
and $\mathcal{B}\,({\tau \to e \gamma})$ are not so tight such as
$4.4~ \times~ 10^{-8}$  and $3.3~ \times~ 10^{-8}$ \cite{BaBar:2009hkt,Belle:2021ysv},
respectively.

On the other hand, theoretical studies 
of the electric and  magnetic dipole moments  of leptons 
are given in the SM Effective Field Theory (SMEFT)
\cite{Buchmuller:1985jz,Grzadkowski:2010es,Alonso:2013hga}, i.e., under the hypothesis of new degrees of freedom
above the electroweak scale \cite{Panico:2018hal,Aebischer:2021uvt,Allwicher:2021rtd,Kley:2021yhn}.
The phenomenological discussion of NP has presented taking  the anomaly of the muon $(g-2)_\mu$ and the LFV bound in the SMEFT.
The flavor symmetry is a challenging hypothesis to reduce the number of independent parameters of the flavor sector.
Indeed, LFV decays and the electron EDM  have been studied 
in the light of the muon $(g-2)_\mu$ anomaly by  imposing 
  $U(2)_L\otimes U(2)_R$ flavor symmetry \cite{Isidori:2021gqe,Tanimoto:2023hse} and some other symmetries~\cite{Calibbi:2021qto}.

Recently, the modular invariance opened up a new promising 
approach to the flavor problem
of quarks and leptons \cite{Feruglio:2017spp}
(see also Refs.\cite{Kobayashi:2018vbk,Penedo:2018nmg,Novichkov:2018nkm}).
Among finite $\Gamma_N$ modular groups, the $\Gamma_3$ modular group, which is {isomorphic} to $A_4$
\cite{Ma:2001dn,Babu:2002dz,Altarelli:2005yp,Altarelli:2005yx,
	Shimizu:2011xg,Petcov:2018snn,Kang:2018txu}, has been extensively used for understanding 
the origin of the quark and lepton flavors
\cite{Feruglio:2017spp,Criado:2018thu,Kobayashi:2018scp,deAnda:2018ecu,Novichkov:2018yse,Okada:2018yrn,Ding:2019zxk,Okada:2019uoy,Okada:2020rjb,Okada:2020ukr,Okada:2020brs,Okada:2021qdf,Kobayashi:2021pav,Kobayashi:2022jvy,Petcov:2022fjf,Petcov:2023vws}. 
Other finite modular groups 
have been also  widely employed in flavor model building 
(see, e.g., \cite{Kobayashi:2018vbk,Penedo:2018nmg,Novichkov:2018ovf,Novichkov:2018nkm,Liu:2019khw,Novichkov:2020eep,Yao:2020zml,Ding:2021iqp,Novichkov:2021evw,Ding:2021zbg,Li:2021buv,deMedeirosVarzielas:2023crv,Ding:2023ydy,
	Meloni:2023aru,Marciano:2024nwm}
and the reviews  
\cite{Altarelli:2010gt,Ishimori:2010au,Ishimori:2012zz,Hernandez:2012ra,King:2013eh,King:2014nza,Tanimoto:2015nfa,King:2017guk,Petcov:2017ggy,Kobayashi:2022moq,Feruglio:2019ybq,Kobayashi:2023zzc}).
Furthermore, the modular symmetry is also  developing to the
strong CP problem \cite{Feruglio:2023uof,Higaki:2024jdk,Petcov:2024vph,Penedo:2024gtb,Feruglio:2024ytl} and modular inflation \cite{Abe:2023ylh,Ding:2024neh,King:2024ssx,Casas:2024jbw}.
It is also  remarked that the formalism of non-holomorphic modular flavor symmetry is developed \cite{Qu:2024rns}.

In the framework of the $A_4$ modular flavor symmetry,
 LFV decays  and the electron EDM  have been studied with the  muon  $(g-2)_{\mu}$ anomaly
 \cite{Kobayashi:2021pav,Kobayashi:2022jvy}.
 In those works,  assuming  NP  to be  heavy and  given by the
SMEFT Lagrangian,  the dipole operator of leptons and their Wilson coefficients  were discussed at the electroweak scale.
Although modular flavor models have been constructed in the supersymmetric framework so far, the modular invariant SMEFT will be 
realized in the so-called moduli-mediated supersymmetry breaking scenario \cite{Kikuchi:2022pkd}.
Furthermore, higher-dimensional operators also keep the modular invariance in a certain class of the string effective field theory 
in which $n$-point couplings of matter fields are written by a product of 3-point couplings \cite{Kobayashi:2021uam}. This  is called
 {\it Stringy Ansatz} to constrain the higher-dimensional operators in the SMEFT.


In this work,  we  take the level 2 finite modular group, 
$\Gamma_2$ \cite{Kobayashi:2018vbk}, which is {isomorphic}  to the 
$S_3$ group,
as the flavor symmetry.
In the $S_3$ group, irreducible representations are two different singlets  $\bf 1$, $\bf 1'$ and one doublet  $\bf 2$,
which are assigned to quarks or leptons.
Similarly,  the light two families
 are assigned to  $\bf 2$ and the third one is $\bf 1$
  in the  $U(2)$ flavor model.
  The $U(2)$ flavor symmetry is successful
   in the quark sector such as  $B$ meson physics
   \cite{Fuentes-Martin:2019mun,Faroughy:2020ina}.
   On the other hand,
the drawback of the $U(2)$ leptonic flavor model
 is  in the difficulty of building a simple  neutrino mass matrix.
 Therefore, the neutrino mass matrix is not specified in the analysis of
 the LFV decays and the electron EDM by  imposing 
 $U(2)$  in Refs.\cite{Isidori:2021gqe,Tanimoto:2023hse}.
 In the $S_3$ flavor model,
 simple lepton mass matrices possibly reproduce the masses and 
 neutrino mixing angles under  the $\Gamma_2$ modular invariance, where the modular symmetry and CP symmetry are broken by fixing the vacuum expectation value (VEV) of modulus $\tau$.
Indeed, the successful models of leptons have been presented \cite{Meloni:2023aru,Marciano:2024nwm}.
Based on  these  works,
 we investigate  LFV decays  and the electron EDM   relating with the  muon  $(g-2)_{\mu}$ anomaly.
 
The paper is organized as follows.
In section \ref{sec:framework}, we present our framework, that is  {\it Stringy Ansatz} and the level 2 modular group.
In section \ref{sec:Wilson}, we show the flavor structure of the Wilson coefficients
of the leptonic dipole operator in mass basis with input data.
In section \ref{sec:S3model}, the $\Gamma_2$ modular invariant models of leptons are introduced.
In section \ref{sec:Wilson-mass}, the Wilson coefficients are obtained approximately in mass basis.
In section \ref{sec:numerical}, we discuss the phenomenology of the  electron $(g-2)_e$, LFV decays and the electron EDM numerically.
Section \ref{sec:summary} is devoted to the summary.
In Appendix \ref{Tensor}, we present  the tensor product of the  $S_3$ group.
In Appendix \ref{exp},  we present the experimental constraints 
on the leptonic dipole operator.
In Appendix  \ref{app-1:UL-UR} and  \ref{app-2:UL-UR}, 
we derive the left-handed and right-handed mixing matrices.
 In Appendix \ref{distribution},
model parameters are discussed in the normal distribution. 

\section{Framework}
\label{sec:framework}
\subsection{Stringy Ansatz}
\label{subsec:string}


{
It was known that $n$-point couplings $y^{(n)}$ of matter fields are written by products of 3-point couplings $y^{(3)}$ in a certain class of string compactifications. 
For instance, 4-point couplings $y^{(4)}_{ijk\ell}$ of matter fields are given by
\begin{equation}
\label{eq:Ansatz}
y^{(4)}_{ijk\ell}=\sum_m y^{(3)}_{ijm}y^{(3)}_{mk\ell}\,,
\end{equation}
up to an overall factor where the subscripts "$\{i,j,k, \ell\}$" indicate states associated with corresponding 4-point interaction
\cite{Kobayashi:2021uam}. 
Here, the virtual modes $"m"$ are light or heavy modes, depending on the compactifications. 
It indicates that the flavor structure of 3-point couplings and higher-dimensional operators has a common 
origin in string compactifications. 
As discussed in Ref. \cite{Kobayashi:2021uam}, such a relation holds at the low-energy scale below the 
compactification scale.

Indeed, the Ansatz Eq.\,(\ref{eq:Ansatz}) is used to predict  NP of leptonic phenomena 
	 in the modular flavor symmetry, as will be discussed in the next sections. 
	}

\subsection{ Modular forms of $\Gamma_2$ modular group}
\label{subsec:level-2}
The $\Gamma_2$ modular group is {isomorphic} to $S_3$
\cite{Kobayashi:2018vbk}, where the irreducible representations are
\begin{align}
{\bf 1}\,,\quad {\bf 1'} \,,\quad {\bf 2}\,.
\end{align}
These tensor products are given in Appendix \ref{Tensor}.
By using 
the Dedekind eta-function $\eta(\tau)$:  
\begin{align}
\eta(\tau) = q^{1/24} \prod_{n =1}^\infty (1-q^n) \,,\qquad
q = e^{2 \pi i \tau}\,,
\end{align}
the modular forms of weight 2 corresponding to the $S_3$ doublet are given as \cite{Kobayashi:2018vbk},
\begin{align}
{\bf Y^{(\rm 2)}_2}
=\begin{pmatrix}Y_1(\tau)\\Y_2(\tau)\end{pmatrix}\,,
\end{align}
where
\begin{eqnarray}
\label{eq:Y-S3}
Y_1(\tau) &=& \frac12 c\left( \frac{\eta'(\tau/2)}{\eta(\tau/2)}  +\frac{\eta'((\tau +1)/2)}{\eta((\tau+1)/2)}
- \frac{8\eta'(2\tau)}{\eta(2\tau)}  \right), \nonumber \\
Y_2(\tau) &=& \frac{\sqrt{3}}{2}c\left( \frac{\eta'(\tau/2)}{\eta(\tau/2)}  -\frac{\eta'((\tau +1)/2)}{\eta((\tau+1)/2)}   \right) \,.\label{doubletY} 
\end{eqnarray}
Here $\tau$ is the complex modulus.
In these expressions,
$c$ is a normalization constant. The   $S_3$ generators
$S$ and $T$ are   in the doublet representation:
\begin{equation}
S = \frac{1}{2}\left(
\begin{array}{cc}
-1        & -\sqrt{3} \\
-\sqrt{3} & 1         
\end{array}\right), \qquad\qquad
T = \left(
\begin{array}{cc}
1 & 0  \\
0 & -1 
\end{array}\right).
\label{S3base}
\end{equation}
Taking $c$ as \cite{Meloni:2023aru}:
\begin{align}
c=i\frac{7}{25\pi}\,,
\end{align}
the doublet modular forms of weight 2 have the following  $q$-expansions:
\begin{align}
{\bf Y^{(\rm 2)}_2}
=\begin{pmatrix}Y_1(\tau)\\Y_2(\tau)\end{pmatrix}
\simeq \begin{pmatrix}  \frac{7}{100}
(1+24q+24q^2+96 q^3+\dots) \\
\frac{14}{25} \sqrt{3}q^{1/2}(1+4q+6q^2+\dots) \end{pmatrix}\,.
\label{Y(2)}
\end{align}
In the basis of Eq.(\ref{S3base}), we can construct modular forms of weight 4 by the tensor product of the two doublets $(Y_1(\tau), Y_2(\tau))^T$:
\begin{align}
{\bf 1}~ & : ~~ {\bf Y^{(4)}_{ 1}} = Y_1(\tau)^2+Y_2(\tau)^2 , \nonumber\\
{\bf 2}~ & : ~~ {\bf Y^{(4)}_{ 2}} =  
\begin{pmatrix}
Y^{(4)}_{1}  \\
Y^{(4)}_{2}
\end{pmatrix}=                                                 
\begin{pmatrix}
Y_2(\tau)^2 - Y_1(\tau)^2  \\
2Y_1(\tau)Y_2(\tau)
\end{pmatrix}.
\label{weight-4}
\end{align}
The $S_3$ singlet ${\bf 1}'$ modular form of the weight 4 vanishes.

Likewise, we obtain the modular forms of weight 6 by tensor products
of three modular forms with weight 2 as:

\begin{align}
{\bf 1}~ & : ~~ {\bf Y^{(6)}_1} = 3Y_1(\tau)Y_2(\tau)^2-Y_1(\tau)^3\,,
\qquad 
{\bf 1'}~  : ~~ {\bf Y^{(6)}_{1'}} = Y_2(\tau)^3-3Y_1(\tau)^2Y_2(\tau)\,, \nonumber\\
{\bf 2}~ & : ~~ {\bf Y^{(6)}_2} =                                                 
\begin{pmatrix}
Y^{(6)}_{1}  \\
Y^{(6)}_{2}
\end{pmatrix}=                                              
\begin{pmatrix}
Y_1(\tau)(Y_1(\tau)^2 + Y_2(\tau)^2)  \\
Y_2(\tau)(Y_1(\tau)^2 + Y_2(\tau)^2)
\end{pmatrix}\, .
\label{weight-6}
\end{align}
In the case of the large ${\rm Im}\, \tau$,
the modular forms of $Y_1$ and $Y_2$
are given {by} $q$ expansions in good approximation.
By using small parameter $\epsilon$,
the  modular forms  are written up to
{${\cal O}(\epsilon)$} as:
\begin{align}
{\bf Y^{(\rm 2)}_2}=\begin{pmatrix}Y_1(\tau)\\Y_2(\tau)\end{pmatrix}
\simeq
\frac{7}{100}
\begin{pmatrix}
1+24\epsilon p\\
8\sqrt{3} \sqrt{\epsilon} p'  (1+4\epsilon p)\end{pmatrix}\,,
\label{app-2}
\end{align}
where
\begin{align}
\epsilon=\exp\,[-2\pi {\rm Im}\,\tau]\,,\qquad
p=\exp\,[2\pi i\, {\rm Re}\,\tau]\,,\qquad 
p'=\exp\,[\pi i\, {\rm Re}\,\tau] \,.
\label{para}
\end{align}
Modular forms of weight 4 and 6 are written
in terms of $\epsilon$, $p$ and $p'$ as:
\begin{align}
&{\bf Y_{\bf 1}^{(4)}}\simeq \left (\frac{7}{100}\right )^2 (1+240 \epsilon p)\,,\quad Y_1^{(4)}\simeq\left (\frac{7}{100}\right )^2 (-1+144 \epsilon p)\, ,\quad Y_2^{(4)}\simeq\left (\frac{7}{100}\right )^2 
(16\sqrt{3} \sqrt{\epsilon} p')\, , \nonumber\\
&{\bf Y_{\bf 1}^{(6)}}\simeq
\left (\frac{7}{100}\right )^3(-1+504  \epsilon  p)\,,
\qquad {\bf Y_{\bf 1'}^{(6)}}\simeq
\left (\frac{7}{100}\right )^3( - 24\sqrt{3} \sqrt{\epsilon } p')\,,\nonumber\\
&Y_1^{(6)}\simeq\left (\frac{7}{100}\right )^3
(1+264 \epsilon p)\, ,
\qquad 
Y_2^{(6)}\simeq\left (\frac{7}{100}\right )^3
(8\sqrt{3}  \sqrt{\epsilon } p')\,  .
\label{app}
\end{align}
\section{Constraints of Wilson coefficients of  dipole operator}
\label{sec:Wilson}

\subsection{Input experimental data}
\label{Input0}

The combined result  
from the E989 experiment at Fermilab ~\cite{Muong-2:2023cdq,Muong-2:2021ojo} and the E821 experiment at BNL~\cite{Muong-2:2006rrc} on $a_\mu=(g-2)_\mu/2$,
together with the SM prediction $a_\mu^\mathrm{SM}$ in~\cite{Aoyama:2020ynm}, implies
\begin{align}
\Delta a_\mu = a_\mu^\mathrm{Exp} - a_\mu^\mathrm{SM} = \brackets{249 \pm 49} \times 10^{-11}~.
\label{muon-data}
\end{align}
We suppose that  $\Delta a_\mu$ comes from NP.

Although
the precise value of the SM prediction of HVP is still  unclear,
 we take a following reference  value
(the discrepancy of  $5.1\,\sigma$ with the SM prediction)  as the input in our numerical analysis:
\begin{align}
\Delta a_\mu &=  249 \times 10^{-11}\,.
\label{muon-input}
\end{align}
We also {impose} the upper-bound  of the absolute value of electron EDM
by the JILA group \cite{Roussy:2022cmp}:
\begin{align}
|{d_e} |<4.1 \times 10^{-30} \, \rm  e\,cm =6.3\times 10^{-14} \,TeV^{-1}\,.
\label{eEDM-input}
\end{align} 
On the other hand, 
the upper-bound of the muon EDM
is \cite{Muong-2:2008ebm}:
\begin{align}
|{d_\mu} |<1.8 \times 10^{-19} \, \rm  e\,cm =2.76\times 10^{-3} \,TeV^{-1}\,.
\label{muEDM}
\end{align} 

The tauon EDM can be evaluated through the measurement of CP-violating correlations in tauon-pair production such as $e^+e^-\to \tau^+\tau^-$
\cite{Belle:2002nla} (see also \cite{Bernreuther:2021elu}).
The present upper-bound on the tauon EDM $d_\tau$ is given as {\cite{Uno:2022xau}:}
\begin{align}
&-1.85\times  10^{-17}{\rm e\,cm}<{\rm Re} \,d_\tau<
0.61\times  10^{-17}{\rm e\,cm}\,,\nonumber\\
&-1.03\times  10^{-17}{\rm e\,cm}<{\rm Im}\, d_\tau<
0.23\times  10^{-17}{\rm e\,cm}\,.
\label{tauEDM0}
\end{align} 
Taking the bound of ${\rm Re} \,d_\tau$, we have 
\begin{align}
|d_\tau|<1.85\times  10^{-17}{\rm e\,cm}= {2.84 \times 10^{-1}}\,{\rm TeV}^{-1}\,.
\label{tauEDM}
\end{align}

The experimental upper-bound for the branching ratio of
the $\mu\to e\gamma$ decay is 
{ \cite{TheMEG:2016wtm,MEGII:2023ltw}:
\begin{align}
\mathcal{B}({\mu^+ \to e^+ \gamma}) < 3.1 \times 10^{-13}\,.
\label{LFV-input}
\end{align} 
}
We also take account of the upper-bound for LFV  decays
 $\tau\to\mu\gamma$  and  $\tau\to e\gamma$ \cite{BaBar:2009hkt,Belle:2021ysv} :
\begin{align}
\mathcal{B}({\tau \to \mu \gamma}) < 4.2 \times 10^{-8}\,,\quad\qquad
\mathcal{B}({\tau \to e \gamma}) < 3.3 \times 10^{-8}\,.
\label{LFV-input2}
\end{align}

These input data are summarized in Table \ref{tab:data}.
	They are converted into the magnitudes of the Wilson coefficients of the  leptonic dipole operator in the next subsection.

\begin{table}[t!]
	\begin{center}
		\renewcommand{\arraystretch}{1.1}
		\begin{tabular}{|c|l|l|} 
			 \hline
			Observables &  Exp.$-$ SM / upper bound & Wilson Coef. in $1/\Lambda^2\ [\mathrm{TeV}^{-2}]$  \\ \hline
			$\Delta a_\mu$ & $249 \times 10^{-11}$\cite{Muong-2:2023cdq,Muong-2:2021ojo,Muong-2:2006rrc} 
			& $\text{Re}\  [\CL{e\gamma}{\mu\mu}^\prime] =1.0 \times 10^{-5} 
			$\\ 
			$\mathcal{B}({\mu^+ \to e^+ \gamma}) $& $< 3.1 \times 10^{-13}$ \cite{TheMEG:2016wtm}
			& $|\CL{e\gamma}{e\mu(\mu e)}^\prime| <1.8 \times 10^{-10}  $\\
			$\mathcal{B}({\tau \to \mu \gamma}) $& $< 4.2 \times 10^{-8}$ \cite{BaBar:2009hkt,Belle:2021ysv}
			& $ | \CL{e\gamma}{\mu \tau(\tau\mu)}^\prime| <2.65 \times 10^{-6}  $\\
			$\mathcal{B}({\tau \to e \gamma}) $&  $< 3.3 \times 10^{-8}$ \cite{BaBar:2009hkt,Belle:2021ysv}
			& $|\CL{e\gamma}{e\tau(\tau e)}^\prime|<  2.35 \times 10^{-6}$ \\ 
			$|d_e| $ & $<4.1 \times 10^{-30} \, \rm  e\,cm$ \cite{Roussy:2022cmp}&$\text{Im}\  [\CL{e\gamma}{ee}^\prime]<1.8\times 10^{-13}$ \\ 
			$|d_\mu| $ & $<1.80 \times 10^{-19} \, \rm  e\,cm$\cite{Muong-2:2008ebm}  &$\text{Im}\  [\CL{e\gamma}{\mu\mu}^\prime]<7.9\times 10^{-3} $ \\ 
			$|d_\tau |$ & $<1.85 \times 10^{-17} \, \rm  e\,cm$ \cite{Belle:2002nla}  &$\text{Im}\  [\CL{e\gamma}{\tau\tau}^\prime]<8.2\times 10^{-1} $ 
			 \\ \hline
		\end{tabular}
	\end{center}
	\caption{Relevant observables and the corresponding values of Wilson coefficients, which
		are presented in $1/\Lambda^2\  (\mathrm{TeV}^{-2})$ unit.}
	\label{tab:data}
\end{table}

\subsection{Wilson coefficients of  leptonic dipole operator}
\label{Wilson-c}
We make the assumption that NP is heavy and can be given by the
SMEFT Lagrangian.
Let us focus on  the dipole operator of leptons and their Wilson coefficients.
The dipole operators come from
SMEFT Lagrangian  at the weak scale as:
\begin{align}
&\mathcal{O}_{\underset{RL}{e\gamma}}
= \frac{v }{\sqrt{2}}  \overline{E}_{R}  \sigma^{\mu\nu} E_{L} F_{\mu\nu}\,,\qquad\qquad
\CL{e\gamma}{RL}^\prime=
\begin{pmatrix}
\CL{e\gamma}{ee}^\prime &\CL{e\gamma}{e\mu}^\prime
&\CL{e\gamma}{e\tau}^\prime\\
\CL{e\gamma}{\mu e}^\prime &\CL{e\gamma}{\mu\mu}^\prime
&\CL{e\gamma}{\mu\tau}^\prime\\
\CL{e\gamma}{\tau e}^\prime &\CL{e\gamma}{\tau\mu}^\prime
&\CL{e\gamma}{\tau\tau}^\prime\\
\end{pmatrix} 
\,,
\nonumber\\
&
\mathcal{O}_{\underset{LR}{e\gamma}}
= \frac{v }{\sqrt{2}}  \overline{E}_{L}  \sigma^{\mu\nu} E_{R} F_{\mu\nu}\,,\qquad\qquad
\CL{e\gamma}{LR}^\prime=\CL{e\gamma}{RL}^{\prime\, \dagger}
\,,
\label{dipole-operators}
\end{align}
where $E_L$ and $E_R$ denote three flavors of the left-handed and
right-handed {charged} leptons, respectively,
and $v$ denotes the vacuum expectation value (VEV) of the Higgs field $H$.
The prime of the Wilson coefficients indicates  the mass-eigenstate basis of the charged leptons.
{(Later, we use the notation of the Wilson coefficients
 without  primes in the  basis of the non-diagonal charged lepton mass matrix.)}
The relevant effective Lagrangian is written as:
\begin{align}
\mathcal{L}_{\rm dipole}=
\frac{1}{\Lambda^2}\,\left (
\CL{e\gamma}{RL}^\prime\mathcal{O}_{\underset{RL}{e\gamma}}
+\CL{e\gamma}{LR}^\prime\mathcal{O}_{\underset{LR}{e\gamma}}
\right )
\,,
\end{align}
where $\Lambda$ is a certain mass scale of NP  in the effective theory.

The operator of  Eq.\eqref{dipole-operators} 
	corresponds to the four-field operator of SMEFT
	$[\overline{E}_{R}  \sigma^{\mu\nu} E_{L} H F_{\mu\nu}]$
	by replacing $v$ with $H$.
	In the viewpoint of  {\it Stringy Ansatz} of Eq.\,(\ref{eq:Ansatz}),  
 the lightest mode $"m"$  corresponds to the Higgs doublet.
	If the mode $"m"$ is only Higgs doublet, the flavor structure of bilinear operator $[\,\bar E_R \sigma^{\mu\nu} E_L\,]$  is 
	exactly the same as the mass matrix.
	Obviously, the bilinear operator matrix is diagonal in the basis for mass eigenstates.
	In this case the LFV processes such as $\mu \to e$, $\tau \to \mu$
	and  $\tau \to e$ never happen. 
	However, additional unknown modes $"m"$  in Eq.\,(\ref{eq:Ansatz}) can cause the flavor violation.
	We discuss such a case in our numerical analysis.

In the following discussions, we take the $\Gamma_2$ modular symmetry for leptons.
Most of modular flavor models are supersymmetric models.
Since we discuss the model below the supersymmetry breaking scale, 
the light modes are exactly the same as the SM with two {Higgs doublet} models.
Note that the modular symmetry is still a symmetry of the low-energy 
	effective action below the supersymmetry breaking scale, 
	as confirmed in the moduli-mediated supersymmetry breaking scenario
 \cite{Kikuchi:2022pkd}.
Here the Wilson coefficients are understood to be evaluated at the weak scale 
\footnote{We  neglect the small effect of running below the weak scale. The one-loop effect is small as seen in~\cite{Buttazzo:2020ibd}.}.
Inputting the value  in Eq.\,\eqref{muon-input},
the $\mu\mu$ component of Wilson coefficients is obtained as\cite{Isidori:2021gqe} 
(see  Appendix \ref{exp}):
\begin{align}
\frac{1}{\Lambda^2}\text{Re}\  [\CL{e\gamma}{\mu\mu}^\prime]
= 1.0 \times 10^{-5} \, \mathrm{TeV}^{-2} \,.
\label{Cmumu-exp}
\end{align}

The LFV process  $\mu \to e \gamma$ gives us more severe constraint
for the $\mu e (e \mu)$ component of Wilson coefficients by the experimental data in Eq.\,\eqref{LFV-input}.
The upper-bound is obtained \cite{Isidori:2021gqe} 
(see Appendix \ref{exp}):
\begin{align}
\frac{1}{\Lambda^2}|\CL{e\gamma}{e\mu(\mu e)}^\prime| <   1.8 \times 10^{-10} \, \mathrm{TeV}^{-2} \, .
\label{eq:bound_C_egamma_12}
\end{align}
Taking into account Eqs.~(\ref{Cmumu-exp}) and \eqref{eq:bound_C_egamma_12}, one has the ratio \cite{Isidori:2021gqe}:
\begin{align}
\left |\frac{ \CL{e\gamma}{e\mu(\mu e)}^\prime  }{    \CL{e\gamma}{\mu\mu}^\prime    }\right | < 1.8 \times 10^{-5}\,.
\label{eq:bound12}
\end{align}
Thus, the magnitude of  $\CL{e\gamma}{e\mu(\mu e)}^\prime $
is much suppressed compared with $\CL{e\gamma}{\mu\mu}^\prime $.
This  gives the severe constraint for parameters of the flavor  model.

The  $\tau \to e \gamma$ and  $\tau \to \mu \gamma$ decays also  give us    constraints
 by the experimental data in Eq.\,\eqref{LFV-input2}.
The upper-bounds of  corresponding components of the Wilson coefficients are obtained as seen in Appendix \ref{exp}:
\begin{align}
\frac{1}{\Lambda^2}|\CL{e\gamma}{\mu\tau(\tau\mu)}^\prime| <  2.65 \times 10^{-6} \, \mathrm{TeV}^{-2} \,, \qquad \qquad
\frac{1}{\Lambda^2}|\CL{e\gamma}{e\tau(\tau e)}^\prime| <  2.35 \times 10^{-6} \, \mathrm{TeV}^{-2}  .
\label{eq:bound_C_egamma_23-13}
\end{align}

The electron EDM,  $d_e$ is defined in the operator:
\begin{align}
\mathcal{O}_{\mathrm{edm}}
= -\frac{i}{2}\,d_e(\mu) \, \overline{e}  \sigma^{\mu\nu} \gamma_5 e F_{\mu\nu}\,,
\label{EDM}
\end{align}
where $d_e=d_e(\mu=m_e)$.
Therefore, the electron EDM is extracted from the effective Lagrangian 
\begin{align}
\mathcal{L}_{\rm EDM}=
\frac{1}{\Lambda^2}\CL{e\gamma}{ee}^\prime \mathcal{O}_{\underset{LR}{e\gamma}} + {\rm h.c.}
= \frac{1}{\Lambda^2}\CL{e\gamma}{ee}^\prime \frac{v }{\sqrt{2}}  \overline{e}_{L}  \sigma^{\mu\nu} e_R F_{\mu\nu}
+ {\rm h.c.}\,,
\label{SMEFT-EDM0}
\end{align}
which leads to 
\begin{align}
d_e=-\sqrt{2} \, \frac{v}{\Lambda^2}\, {\rm Im}\, [\CL{e\gamma}{ee}^\prime] \,,
\label{SMEFT-EDM}
\end{align}
at tree level, where the small effect of running below the electroweak scale is neglected.

Inputting the experimental  upper-bound of the electron EDM
in Eq.\,\eqref{eEDM-input} \cite{Roussy:2022cmp},
we obtain $ee$ component of the constraints of the Wilson coefficient:
\begin{equation}
\frac{1}{\Lambda^2}{\mathrm{Im}}\, [\CL{e\gamma}{ee}^\prime]<1.8\times 10^{-13}\, 
\mathrm {TeV^{-2}}\,.
\label{Cee-exp}
\end{equation}
By taking the value in  Eq.\eqref{Cmumu-exp}, we have
a very small ratio:
\begin{align}
\frac{ {\rm Im}\,[\CL{e\gamma}{e\mu(\mu e)}^\prime]  }
{ {\rm Re}\,[\CL{e\gamma}{\mu\mu}^\prime]    } <  1.8\times 10^{-8}\,.
\label{eq:edm}
\end{align}

On the other hand,  the experimental  upper-bound of the muon EDM
in Eq.\,\eqref{muEDM}  gives:
\begin{equation}
\frac{1}{\Lambda^2}{\mathrm{Im}}\, [\CL{e\gamma}{\mu \mu}^\prime]<7.9\times 10^{-3}\, 
\mathrm {TeV^{-2}}\,.
\end{equation}
The  upper-bound of the tauon EDM
in Eq.\,\eqref{tauEDM} also gives:
\begin{equation}
\frac{1}{\Lambda^2}{\mathrm{Im}}\, [\CL{e\gamma}{\tau \tau}^\prime]<8.2\times 10^{-1}\,  \mathrm {TeV^{-2}}\,.
\end{equation}

These upper-bounds  of the Wilson coefficients are summarized 
	in Table \ref{tab:data}.
Using these upper-bounds of Wilson coefficients, we discuss
NP based on  the $\Gamma_2$ modular invariant flavor model in the next section.


\section{$\Gamma_2$ modular invariant model}
\label{sec:S3model}
\subsection{Mass matrices of leptons}

 In the $\Gamma_2$ modular invariant model, the successful lepton mass matrices are presented in Ref.\cite{Meloni:2023aru}.
 There are two-type charge lepton mass matrices.
 The first one is the minimal model I,
 where the weight 2 and 4 modular forms used, and the second one
 is the hierarchical model {I\hskip -0.4mm I},
 where the weight 2 and 6 modular forms appear.
 
 Let us discuss Model I,
 in which the assignments of  the weights 
 for the relevant chiral superfields as
 in Table \ref{tab:model-I}.
  \begin{table}[H]
 	\begin{center}
 		\renewcommand{\arraystretch}{1.1}
 		\begin{tabular}{|c|c|c|c|c|} \hline
 			& $(e,\mu)_L,\ \ \tau_L$ & $e^c,\ \ \mu^c,\ \ \tau^c$ &  $H_u$ & $H_d$ \\ \hline
 			$SU(2)$ & 2 & 1  & 2 & 2 \\
 			$S_3$ & $2\qquad\ \ 1'$ & $1\ \ \ 1' \ \ \ 1'$ & $1$ & $1$ \\
 			$k$ &$2$  &$2\quad\, 0\   -2$  & 0 & 0 \\ \hline
 		\end{tabular}
 	\end{center}
 	\caption{Assignments of $S_3$ representations and weights in Model I.}
 	\label{tab:model-I}
 \end{table}

The superpotential terms of the  charged lepton masses
 $ w_{E(\rm I)}$
are written as:
  \begin{align} w_{E(\rm I)}=
& \alpha_1 e^c H_d [(e,\mu)_L {\bf Y_2^{\rm (4)}}]_{\bf 1}
+\beta_1 \mu^c H_d [(e,\mu)_L {\bf Y_2^{\rm(2)}}]_{\bf 1'}
+\gamma_1 \tau^c H_d \tau_L  \,.
\label{w-1}
\end{align}
The charged lepton mass matrix is given as:
 \begin{align}
{\rm I :}  \quad  M_E =v_d
 \begin{pmatrix}
 \alpha_1  Y^{(4)}_1 & \alpha_1  Y^{(4)}_2 &0 \\
  \beta_1  Y_2 & -\beta_1  Y_1 &0 \\
 0&0&\gamma_1
 \end{pmatrix}_{RL}\,,
 \label{model-I0}
 \end{align}
 where  $\alpha_1$, $\beta_1$ and  $\gamma_1$ are taken to be 
 real positive parameters without loss of generality, and
 $v_d$ denotes the VEV of $H_d$.
 It  is given approximately
  by using Eqs.\,\eqref{app-2} and \eqref{app} as:
  \begin{align}
{\rm I :} \quad  M_E \simeq v_d
 \begin{pmatrix}
 -\tilde \alpha_1 (1-144 \epsilon p) 
 & 16\sqrt{3}\tilde \alpha_1  \sqrt{\epsilon}p' &0 \\
  8\sqrt{3}\tilde \beta_1 \sqrt{\epsilon}p'& -\tilde\beta_1 (1+24 \epsilon p) &0 \\
 0&0&\gamma_1
 \end{pmatrix}_{RL}\,,
 \label{model-I}
 \end{align}
 where
 \begin{align}
\tilde\alpha_1=\left (\frac{7}{100}\right )^2 \alpha_1\,,
\qquad \tilde\beta_1=\left (\frac{7}{100}\right ) \beta_1\,,
 \end{align}
 while $\gamma_1$ is remained.

  Next, we show Model {I\hskip -0.4mm I},
 in which the assignments of  the weights 
 for the relevant chiral superfields as
 in Table \ref{tab:model-II}.
 \begin{table}[H]
 	\begin{center}
 		\renewcommand{\arraystretch}{1.1}
 		\begin{tabular}{|c|c|c|c|c|} \hline
 			& $(e,\mu)_L,\ \ \tau_L$ & $e^c,\ \ \mu^c,\ \ \tau^c$ &  $H_u$ & $H_d$ \\ \hline
 			$SU(2)$ & 2 & 1  & 2 & 2 \\
 			$S_3$ & $2\qquad\ \ 1'$ & $1\ \ \ 1' \ \ \ 1'$ & $1$ & $1$ \\
 			$k$ &$4$  &$2\quad\, 0\   -2$  & 0 & 0 \\ \hline
 		\end{tabular}
 	\end{center}
 	\caption{Assignments of $S_3$ representations and weights in 
 		Model {I\hskip -0.4mm I}.}
 	\label{tab:model-II}
 \end{table}
The superpotential terms of the charged lepton masses  
$w_{E(\rm {I\hskip -0.4mm I})}$ are given 
  \begin{align} w_{E(\rm {I\hskip -0.4mm I})}=
 \alpha_2 e^c H_d [(e,\mu)_L {\bf Y_2^{\rm (6)}}]_{\bf 1}
+\beta_2 \mu^c H_d [(e,\mu)_L {\bf Y_2^{\rm(2)}}]_{\bf 1'}
+\gamma_2 \tau^c H_d \tau_L 
+ \alpha_D e^c H_d \tau_L {\bf Y_{1'}^{\rm (6)}}
  \,.
\label{w-2}
\end{align}
 The charged lepton mass matrix is given in terms of 
 $\bf 1'$ modular form with weigh 6 in addition to 
 $S_3$ doublet modular forms
 with weight 2 and 6 as:
 \begin{align}
 {\rm {I\hskip -0.4mm I} :} \quad M_E =v_d
 \begin{pmatrix}
 \alpha_2  Y^{(6)}_1 & \alpha_2  Y^{(6)}_2 & \alpha_D {\bf Y^{(\rm 6)}_{1'}} \\
 \beta_2  Y_2 & -\beta_2  Y_1 &0 \\
 0&0&\gamma_2
 \end{pmatrix}_{RL}\,,
 \label{model-II0}
 \end{align}
 where $\alpha_2$, $\alpha_D$, $\beta_2$ and  $\gamma_2$ are
 real positive parameters without loss of generality.
It  is given approximately  as:
 \begin{align}
{\rm {I\hskip -0.4mm I} :} \quad M_E \simeq v_d
\begin{pmatrix}
\tilde\alpha_2 (1+264 \epsilon p) & 8\sqrt{3}\tilde\alpha_2  \sqrt{\epsilon}p'
 &  -24\sqrt{3}\tilde\alpha_D \sqrt{\epsilon}p'\\
 8\sqrt{3}\tilde\beta_2   \sqrt{\epsilon}p' & -\tilde \beta_2 (1+24 \epsilon p) &0 \\
0&0&\gamma_2
\end{pmatrix}_{RL}\,,
\label{model-II}
\end{align}
 where
\begin{align}
\tilde\alpha_2=\left (\frac{7}{100}\right )^3 \alpha_2\,,
\qquad \tilde\alpha_D=\left (\frac{7}{100}\right )^3 \alpha_D\,,
\qquad \tilde\beta_2=\left (\frac{7}{100}\right ) \beta_2\,,
\end{align}
while   $\gamma_2$ is remained.

On the other hand, the Weinberg operator gives the neutrino mass matrix,
which is common for Model I and Model {I\hskip -0.4mm I}:
\begin{align}
M_\nu=2g\frac{v_u^2}{\Lambda}\left [
\begin{pmatrix}
Y_1^2-Y_2^2&2Y_1 Y_2 & 2\frac{g'}{2g}Y_1 Y_2\\
2Y_1 Y_2  & Y_2^2-Y_1^2 &\frac{g'}{2g}(Y_1^2-Y_2^2) \\
2\frac{g'}{2g}Y_1 Y_2 & \frac{g'}{2g}(Y_1^2-Y_2^2) & 0
\end{pmatrix}
+
(Y_1^2+Y_2^2)
\begin{pmatrix}
\frac{g''}{g} & 0 &0\\ 0 &\frac{g''}{g} & 0 \\
 0 & 0 &\frac{g_p}{g}
\end{pmatrix}
\right ] \ ,
\label{NeutrinotauST}
\end{align}
where the $g'$, $g''$ and $g_p$ are real positive parameters.
It is noted that  CP is violated  
 by fixing the modulus $\tau$ since the imaginary part of the lepton mass matrices appear through ${\rm Re}\,\tau$.
 
  However, we do not discuss details of the  neutrino mass matrix in this work
  because its contribution to our result is negligibly small due to small neutrino masses.

\subsection{Wilson coefficients of the leptonic dipole operator}

The leptonic dipole operator is written in the  flavor space 
as seen in Eq.\,\eqref{dipole-operators},
where the charged lepton mass matrix is non-diagonal, and given in
Eqs.\,\eqref{model-I0} or \eqref{model-II0}.

 In Model I,
the flavor structure of $\CL{e\gamma}{RL}$ is {the} same as the Yukawa couplings $Y_{RL}$ apart from coefficients
 $\alpha'_1$,  $\beta'_1$ and  $\gamma'_1$.
 It  is given approximately
by using Eq.\,\eqref{app} as:
\begin{align}
{\rm I :} \quad \CL{e\gamma}{RL}
=\begin{pmatrix}
\CL{e\gamma}{ee} &\CL{e\gamma}{e\mu}
&\CL{e\gamma}{e\tau}\\
\CL{e\gamma}{\mu e} &\CL{e\gamma}{\mu\mu}
&\CL{e\gamma}{\mu\tau}\\
\CL{e\gamma}{\tau e} &\CL{e\gamma}{\tau\mu}
&\CL{e\gamma}{\tau\tau}\\
\end{pmatrix}  \simeq 
\begin{pmatrix}
-\tilde \alpha'_1 (1-144 \epsilon p) 
& 16\sqrt{3}\tilde \alpha'_1  \sqrt{\epsilon}p' &0 \\
 8\sqrt{3}\tilde \beta'_1 \sqrt{\epsilon}p'& -\tilde\beta'_1 (1+24 \epsilon p) &0 \\
0&0&\gamma'_1
\end{pmatrix}_{RL}\,,
\label{C-model-I}
\end{align}
where
\begin{align}
\tilde\alpha'_1=\left (\frac{7}{100}\right )^2 \alpha'_1\,,
\qquad \tilde\beta'_1=\left (\frac{7}{100}\right ) \beta'_1\,.
\end{align}
Coefficients  $\alpha'_1$,  $\beta'_1$ and  $\gamma'_1$ are different from
 $\alpha_1$, $\beta_1$  and  $\gamma_1$ in Eq.\,\eqref{model-I0}.
 Those are  complex parameters 
 in constrast to real positive $\alpha_1$, $\beta_1$ and  $\gamma_1$ . 


In Model {I\hskip -0.4mm I},  it is given approximately  as:
\begin{align}
{\rm {I\hskip -0.4mm I} :} \quad \CL{e\gamma}{RL} =
\begin{pmatrix}
\CL{e\gamma}{ee} &\CL{e\gamma}{e\mu}
&\CL{e\gamma}{e\tau}\\
\CL{e\gamma}{\mu e} &\CL{e\gamma}{\mu\mu}
&\CL{e\gamma}{\mu\tau}\\
\CL{e\gamma}{\tau e} &\CL{e\gamma}{\tau\mu}
&\CL{e\gamma}{\tau\tau}\\
\end{pmatrix} \simeq 
\begin{pmatrix}
{\tilde\alpha'_2} (1+264 \epsilon p) & 8\sqrt{3} {\tilde\alpha'_2}  \sqrt{\epsilon}p'
& -24\sqrt{3} {\tilde\alpha'_D} \sqrt{\epsilon}p'\\
8\sqrt{3} {\tilde\beta'_2}   \sqrt{\epsilon}p'
 & -{\tilde\beta'_2} (1+24 \epsilon p) &0 \\
0&0&\gamma'_2
\end{pmatrix}_{RL}\,,
\label{C-model-II}
\end{align}
where
\begin{align}
\tilde\alpha'_2=\left (\frac{7}{100}\right )^3 \alpha'_2\,,
\qquad \tilde\alpha'_D=\left (\frac{7}{100}\right )^3 \alpha'_D\,,
\qquad \tilde\beta'_2=\left (\frac{7}{100}\right ) \beta'_2\,.
\end{align}
Coefficients  $\alpha'_2$, $\beta'_2$, $\alpha'_D$ and $\gamma'_2$   are different from
$\alpha_2$, $\beta_2$, $\alpha_D$ and   $\gamma_2$ in Eq.\,\eqref{model-II0}
and complex parameters in general. 

The mass matrix $M_E$ in Eqs.\,\eqref{model-I} and \eqref{model-II}
is diagonalized by the bi-unitary transformation
$U_R^\dagger M_E U_L$.
We can obtain $U_L$ and $U_R$ by diagonalizing
$U_L^\dagger M_E^\dagger M_E U_L$ and  $U_R^\dagger M_E M_E^\dagger U_R$,
respectively.
 Finally, we obtain 
    $\CL{e\gamma}{RL}^\prime$ 
 in the real diagonal basis of the charged lepton mass matrix as
 $\CL{e\gamma}{RL}^\prime=U_R^\dagger\  \CL{e\gamma}{RL}   U_L$
 for both Model I and Model {I\hskip -0.4mm I}.
Approximate matrices of $U_L$ and $U_R$ are given 
in Appendix  \ref{app-1:UL-UR} and \ref{app-2:UL-UR}
for Model I and Model {I\hskip -0.4mm I}, respectively.

\section{Wilson coefficients in mass basis}
\label{sec:Wilson-mass}
\subsection{Wilson coefficients in Model I}
\label{W1}
In the flavor basis, the Wilson coefficients are given
in Eq.\,\eqref{C-model-I} for Model I.
We move to the basis of the diagonal mass matrix of the charged leptons
as follows:
	\begin{align}
	\CL{e\gamma}{RL}^\prime \simeq 
 U_{R1}^T P_{R1}^* 
	\begin{pmatrix}
	-\tilde \alpha'_1 (1-144 \epsilon p) 
	&16\sqrt{3}\tilde \alpha'_1  \sqrt{\epsilon}p' &0 \\
	8\sqrt{3}\tilde \beta'_1 \sqrt{\epsilon}p'
	& -\tilde\beta'_1 (1+24 \epsilon p) &0 \\
	0&0&\gamma'_1
	\end{pmatrix} 
	P_{L1} U_{L1}\,,
	\label{app:Cprime-I}
	\end{align}
where  $P_{L1}$, $U_{L1}$, $P_{R1}$ and $ U_{R1}$,
are given in Eqs.\,\eqref{PL1}, \eqref{UL1}, \eqref{PR1} and \eqref{UR1}
of Appendix \ref{app-1:UL-UR}, respectively.
 Here, we take
  $\tilde \alpha_1\ll \tilde \beta_1$
and  $|\tilde \alpha'_1|\ll |\tilde \beta'_1|$.

We obtain the Wilson coefficients up to
${\cal O}(\sqrt{\epsilon})$ as follows::
\begin{align}
&\CL{e\gamma}{ee}^\prime \simeq  
-\tilde \alpha'_1 \, ,
\qquad 
\CL{e\gamma}{\mu\mu}^\prime \simeq 
-\tilde \beta'_1
\,,
\qquad \CL{e\gamma}{ \tau \tau}^\prime=\gamma'_1 \,,
\nonumber\\
&\CL{e\gamma}{ e\mu }^\prime\simeq 
8\sqrt{3}\tilde \alpha'_1 \sqrt{\epsilon}
 e^{-i(\phi_R+\pi\tau_R)} 
\left [1+2e^{2i\pi\tau_R}
-\frac{\tilde \alpha_1}{\tilde \alpha'_1}
\frac{\tilde \beta'_1}{\tilde \beta_1}
e^{i(\phi_R+\pi\tau_R)}\sqrt{5+4\cos 2\pi\tau_R} \right ]\nonumber\\
&\quad \ =8\sqrt{3}\tilde \alpha'_1 \sqrt{\epsilon}
\sqrt{5+4\cos 2\pi\tau_R}
\left (1- \frac{\tilde \alpha_1}{\tilde \alpha'_1}
\frac{\tilde \beta'_1}{\tilde \beta_1}\right )\,,
\nonumber\\
&\CL{e\gamma}{ \mu e}^\prime  \simeq   8\sqrt{3}\tilde \alpha'_1 \frac{\tilde \alpha_1}{\tilde \beta_1}\sqrt{\epsilon}
\left[-\frac{\tilde \alpha_1}{\tilde \alpha'_1}
\frac{\tilde \beta'_1}{\tilde \beta_1}(1+2 e^{-2i\pi\tau_R})e^{i(\phi_R+\pi\tau_R)}+\sqrt{5+4\cos 2\pi\tau_R}\right ]
\nonumber\\
&\quad \ =8\sqrt{3}\tilde \alpha'_1 \frac{\tilde \alpha_1}{\tilde \beta_1}\sqrt{\epsilon}
\sqrt{5+4\cos 2\pi\tau_R}
\left (1- \frac{\tilde \alpha_1}{\tilde \alpha'_1}
\frac{\tilde \beta'_1}{\tilde \beta_1}\right )\,,
\nonumber\\
&\CL{e\gamma}{ e \tau}^\prime=\CL{e\gamma}{ \tau e}^\prime=0\,, \qquad \CL{e\gamma}{ \mu \tau}^\prime=\CL{e\gamma}{ \tau\mu}^\prime=0\,.
\label{Wilson12}
\end{align}
It is noted that
 $\CL{e\gamma}{ e\mu}^\prime$ and  $\CL{e\gamma}{ \mu e}^\prime$
  vanish  if  $\tilde\alpha'_1=\tilde\alpha_1$ and
    $\tilde \beta'_1=\tilde\beta_1$ are put.
   In this model, $e$-$\tau$ and $\mu$-$\tau$ transitions  never occur.
    It is also  remarked 
  that the imaginary part of Wilson coefficients vanish in full orders of $\epsilon$ if   $\tilde\alpha'_1$, $\tilde \beta'_1$ 
  and  $\tilde\gamma'_1$ are real.
  That is, the imaginary parts of $\tilde \alpha'_1$, $\tilde\beta'_1$
  and $\tilde\gamma'_1$ are the origin of the leptonic EDM.
  
  Taking a constraint of Eq.\,\eqref{eq:edm},
  we have 
\begin{align}
 \left | \frac{{\text{Im}\,\tilde \alpha'_1}}{{\text{Re}\,\tilde \beta'_1}} \right |
  < 1.8\times 10^{-8}\,.
\end{align}
Thus, the imaginary part of $\alpha'_1$ should be tiny.
This tiny imaginary part is discussed in the standpoint of 
 {\it Stringy Anzatz} in the next section.

\subsection{Wilson coefficients in Model {I\hskip -0.4 mm I}}
\label{W2}
In the Model {I\hskip -0.4 mm I},   the Wilson coefficients are given
in Eq.\eqref{C-model-II}.
In the basis 
of the diagonal mass matrix of the charged leptons, we have:
as follows:
	\begin{align}
	\CL{e\gamma}{RL}^\prime \simeq v_d
U_{R2}^T P_{R2}^*
	\begin{pmatrix}
	{\tilde\alpha'_2} (1+264 \epsilon p) & 8\sqrt{3} {\tilde\alpha'_2}  \sqrt{\epsilon}p'
	& -24\sqrt{3} {\tilde\alpha'_D} \sqrt{\epsilon}p'\\
8\sqrt{3} {\tilde\beta'_2}   \sqrt{\epsilon}p'
	& - {\tilde\beta'_2} (1+24 \epsilon p) &0 \\
	0&0&\gamma'_2
	\end{pmatrix}
	P_{L2} U_{L2}\,,
	\label{app:Cprime-II}
	\end{align}
	where  $P_{L2}$, $U_{L2}$, $P_{R2}$ and $ U_{R2}$,
	are given in Eqs.\,\eqref{PL2}, \eqref{UL2}, \eqref{PR2} and \eqref{UR2}
	of Appendix \ref{app-2:UL-UR}, respectively.
	Here,  we take $\tilde \alpha_2\sim
	\tilde \alpha_D\ll \tilde \beta_2\ll \gamma_2$ and 
 $|\tilde \alpha'_2|\sim
|\tilde \alpha'_D|\ll |\tilde \beta'_2|\ll |\tilde \gamma'_2|$.

The Wilson coefficients are given explicitly  up to
${\cal O}(\sqrt{\epsilon})$ as follows:
\begin{align}
&\CL{e\gamma}{ee}^\prime \simeq  
\tilde \alpha'_2 \, , \qquad
\CL{e\gamma}{\mu\mu}^\prime \simeq 
-\tilde \beta'_2\,, \qquad\CL{e\gamma}{\tau\tau}^\prime \simeq 
\gamma'_2\,,
\nonumber\\
&\CL{e\gamma}{ e\mu }^\prime\simeq 
8\sqrt{3}\tilde \alpha'_2 \sqrt{\epsilon}\,
i e^{-i\pi\tau_R}
\left (-1+e^{2i\pi\tau_R}-
2i\,\frac{\alpha_2}{\alpha'_2}\frac{\beta'_2}{\beta_2}e^{i\pi\tau_R}
\sin\pi\tau_R\right )\nonumber\\
&\quad\ =-16\sqrt{3}\tilde \alpha'_2 \sqrt{\epsilon}\,  
\sin\pi\tau_R
\left(1-\frac{\alpha_2}{\alpha'_2}\frac{\beta'_2}{\beta_2}  \right ) \,,
\nonumber\\
&{\CL{e\gamma}{ \mu e}^\prime  \simeq 
8\, i\,e^{i\pi\tau_R}\sqrt{3}\tilde \beta'_2 \frac{\tilde \alpha^2_2}{\tilde \beta^2_2}\sqrt{\epsilon}
\left(-1+e^{-2i\pi\tau_R}+
2i\frac{\alpha'_2}{\alpha_2}\frac{\beta_2}{\beta'_2} e^{-i\pi\tau_R}\sin\pi\tau_R \right ) 
}\nonumber\\
&{\qquad\ =16\sqrt{3}\tilde \beta'_2 \frac{\tilde \alpha^2_2}{\tilde \beta^2_2}\sin\pi\tau_R \sqrt{\epsilon}
\left(1-\frac{\alpha'_2}{\alpha_2}\frac{\beta_2}{\beta'_2} \right ) 
}\,,\nonumber\\
&\CL{e\gamma}{ e\tau}^\prime \simeq -24\sqrt{3}\tilde \alpha'_D
\sqrt{\epsilon} \left (1-\frac{\tilde\alpha_D}{\tilde \alpha'_D} \frac{\gamma'_2}{\gamma_2}\right )\,,
\nonumber\\
&\CL{e\gamma}{ \tau e}^\prime \ \simeq
	-24\sqrt{3} \alpha'_2 \frac{\tilde\alpha_D}{\gamma_2}\sqrt{\epsilon}
\left (1-\frac{\tilde\alpha_2}{\tilde \alpha'_2} \frac{\gamma'_2}{\gamma_2}\right )	
\,,
\nonumber\\
&\CL{e\gamma}{ \mu \tau}^\prime \sim
\tilde\alpha'_D\, {\cal O} \left(\frac{\tilde \alpha_2}{ \tilde\beta_2}\epsilon \right) \,,
\qquad 
\CL{e\gamma}{  \tau\mu}^\prime \sim \tilde \alpha'_2\, {\cal O} \left(\frac{\tilde \alpha_D}{ \gamma_2}\epsilon \right) \,.
\label{Wilson22}
\end{align}
It is easily found that 
$\CL{e\gamma}{ e\mu}^\prime$ and  $\CL{e\gamma}{ \mu e}^\prime$
vanish  if  $\tilde\alpha'_2=\tilde\alpha_2$ and
$\tilde \beta'_2=\tilde\beta_2$ are put.
The coefficient $\CL{e\gamma}{ e\tau}^\prime$ 
($\CL{e\gamma}{ \tau e}^\prime$) also vanishes 
 if  $\tilde\alpha'_D=\tilde\alpha_D$ ($\tilde\alpha'_2=\tilde\alpha_2$) and
$\tilde \gamma'_2=\tilde\gamma_2$ are {impose}.
Since other coefficients  $\CL{e\gamma}{\teal{ \tau e} }^\prime$,
  $\CL{e\gamma}{ \mu\tau}^\prime$ and 
   $\CL{e\gamma}{ \tau\mu}^\prime$ are suppressed,
   we present them in order estimates.

It is also  remarked 
that the imaginary part of Wilson coefficients vanish in full orders of $\epsilon$ if   $\tilde\alpha'_2$, $\tilde \beta'_2$ 
and $\tilde\gamma'_2$ are real.
That is, the imaginary part of $\tilde \alpha'_2$, $\tilde\beta'_2$
and $\tilde\gamma'_2$ lead to the leptonic EDM.

Taking a constraint of Eq.\,\eqref{eq:edm},
 the imaginary part of $\alpha'_2$ should be also  tiny
  as discussed in subsection \ref{W1}.

\section{Numerical analyses}
\label{sec:numerical}
\subsection{Parametrization}
As discussed in subsection \ref{Wilson-c},
	the lightest mode $"m"$  corresponds to Higgs doublet $H$
	in    {\it Stringy Ansatz} of Eq.\,(\ref{eq:Ansatz}).
	If the mode $"m"$ is only $H$, the flavor structure of bilinear operator $[\,\bar E_R \sigma^{\mu\nu} E_L\,]$  is 
	exactly the same as the mass matrix.
	Therefore,
	the LFV process such as $\mu \to e$ never happen. 
	However, additional unknown modes $"m"$ in Eq.\,(\ref{eq:Ansatz}) cause the flavor violation.
	We discuss such a case in our numerical analysis.
	Let us introduce a small parameter $\delta_\alpha$
    to see the difference between
	$\tilde\alpha_{1(2)}$ and $\tilde\alpha'_{1(2)}$.
  {In the same way},  $\delta_\beta$, $\delta_\gamma$ and $\delta_D$ are
  introduced. They are expected to be of the same order
   and given as follows:
	\begin{eqnarray}
	 \frac{\tilde\alpha'_{1(2)}}{\tilde \alpha_{1(2)}}= 1+\delta_\alpha\, ,\qquad 
	 \frac{\tilde\beta'_{1(2)}}{\tilde \beta_{1(2)}}= 1+\delta_\beta\, ,
	 \qquad 
	 \frac{\tilde\gamma'_{1(2)}}{\tilde \gamma_{1(2)}}= 1+\delta_\gamma\, ,
	 \qquad 
	 \frac{\tilde\alpha'_{D}}{\tilde \alpha_{D}}= 1+\delta_D\,,
	\label{deviation}
	\end{eqnarray}
	where $\delta_\alpha$, $\delta_\beta$, $\delta_\gamma$ and $\delta_D$  are complex and tiny   from   unknown modes of $"m"$ in Eq.\,\eqref{eq:Ansatz},
	which may be higher excited modes of Higgs. 
	Indeed, if $\delta_\alpha$, $\delta_\beta$, $\delta_\gamma$ and $\delta_D$ vanish,   off diagonal
	Wilson coefficients vanishes as seen in Eqs.\,\eqref{Wilson12} and \eqref{Wilson22}
	of Section \ref{sec:Wilson-mass}
	up to	${\cal O}(\sqrt{\epsilon})$. 
We constrain those small parameters   by inputting the experimental
	upper-bounds in subsection \ref{Input0}.

{In the following analyses,  small parameters
	$\delta_\alpha$ et al. are put statistically in
	the normal distribution with an average $0$ and standard deviation 
	$\sigma$.  Then,  we can take 
	$|\delta_\alpha|\simeq|\delta_\beta|\simeq|\delta_\gamma|\simeq
	|\delta_D|\simeq \sigma$ as seen in Appendix \ref{distribution}.
}
\subsection{Input parameter}
\label{Input}
Since we adopt the $\Gamma_2$ modular flavor models in Ref.\cite{Meloni:2023aru}, the parameters of charged lepton mass matrix
are fixed.
We list them for  the charged lepton sector in Table \ref{tab:parameters} as follows:
\begin{table}[H]
	\begin{center}
		\renewcommand{\arraystretch}{1.1}
		\begin{tabular}{|c|c|c|c|} \hline
			\rule[14pt]{0pt}{2pt} 	
			& Model I & Model I\hskip-0.4mm I & $''$Seesaw Model$''$  \\ \hline
			\rule[14pt]{0pt}{2pt} 	
				$\tau$	&$\pm 0.0895+1.697\,i$ &$\pm 0.090+1.688\,i$ & $\pm 0.244+1.132\,i$  \\ \hline
					\rule[14pt]{0pt}{2pt} 	
				$\beta/\alpha$	&$14.33$ &$1.03$ & $0.92$  \\ \hline
					\rule[14pt]{0pt}{2pt} 	
				$\gamma/\alpha$	&$17.39$ &$1.26$ & $-1.20$  \\ \hline
					\rule[14pt]{0pt}{2pt} 	
				$\alpha_D/\alpha$	&$-$ &$1.33$ & $10^{-13.4}$  \\ \hline
		\end{tabular}
	\end{center}
	\caption{Best fit values of parameters of the charged leptons in $\Gamma_2$ modular flavor model
		 \cite{Meloni:2023aru},
	 where $\alpha$,  $\beta$ and $\gamma$
 correspond to $\alpha_1$,  $\beta_1$ and  $\gamma_1$
  for Model I,  and $\alpha_2$,  $\beta_2$ and  $\gamma_2$ for Model I\hskip-0.4mm I, respectively.
Values in
 $''$Seesaw Model$''$ are also listed \cite{Marciano:2024nwm}.}
 	\label{tab:parameters}
\end{table}
We have not discussed the "Seesaw Model" 
in Table \ref{tab:parameters} since the charged lepton mass matrix is the same flavor structure as Model {I\hskip-0.4mm I}.
 We only comment on the numerical result for "Seesaw Model" 
 in Summary of section \ref{sec:summary}.
 We use these values in our numerical calculations.

{
\subsection{Constraints of $\delta_{\alpha}$ and  $\delta_{\beta}$
	from $|d_e|$ and  $\mathcal{B}({\mu \to e \gamma})$}
\label{constraint-delta}
The tight constraints come from the experimental upper-bounds of  $|d_e|$ and  $\mathcal{B}({\mu \to e \gamma})$.
The relevant  Wilson coefficients 
$\CL{e\gamma}{ ee}^\prime$ and  $\CL{e\gamma}{e \mu}^\prime$
are given in Eqs.\,\eqref{Wilson12} and \eqref{Wilson22}.
Let us consider the  Model I to see the magnitude of constraint from  $|d_e|$ and  $\mathcal{B}({\mu \to e \gamma})$.
By using the approximate forms in 
 Eq.\,\eqref{sigma} of Appendix \ref{distribution},
 we have a simple expressions of the ratio of the  Wilson coefficients:
 \begin{align}
 \left |\frac{ \CL{e\gamma}{e\mu}^\prime  }{    \CL{e\gamma}{\mu\mu}^\prime    }\right | 
  \simeq 
  8\sqrt{3}\sqrt{\epsilon}\sqrt{5+4\cos 2\pi\tau_R}\
 \left |\frac{\tilde\alpha'_1}{\tilde\beta'_1}  \left ( 1-\frac{\tilde{\alpha_1}}{\tilde{\alpha'_1}}
  \frac{\tilde{\beta'_1}}{\tilde{\beta_1}} \right )\right |
 \simeq 
  8\sqrt{3}\sqrt{\epsilon}\sqrt{5+4\cos 2\pi\tau_R}\left |\frac{\tilde\alpha'_1}{\tilde\beta'_1} \right |\, \sigma.
 \end{align}
 Putting the numerical values of Model I in Table \ref{tab:parameters},
 we have
 \begin{align}
 \left |\frac{ \CL{e\gamma}{e\mu(\mu e)}^\prime  }{    \CL{e\gamma}{\mu\mu}^\prime    }\right |\simeq 10^{-3} \sigma\,,
 \end{align}
 where $|\tilde\alpha'_1/\tilde\beta'_1|\simeq (7/100)\times \alpha_1/\beta_1$ and $\sqrt{\epsilon}=5\times 10^{-3}$ are used.
 Since the upper-bound of this  {ratio} is {$1.8 \times 10^{-5}$} as seen in  Eq.\,\eqref{eq:bound12}, we have 
 \begin{align}
 \sigma\simeq |\delta_\alpha|\simeq |\delta_\beta|< 10^{-2}\,.
 \label{Abs-delta}
 \end{align}
 Indeed,  in our numerical calculations,
 the statistical  parameters $\delta_\alpha$
  and  $\delta_\beta$ with the average $0$ and  $\sigma=0.01$
  reproduce 
   $\mathcal{B}({\mu \to e \gamma})$ consistent with the experimental upper-bound. 
   
   It is remarked  that the NP signal of the $\mu\to e\gamma$ process comes from the  operator $\bar e_R \sigma_{\mu\nu}\mu_L$ mainly
   in our scheme because we have a ratio 
    \begin{align}
   \left |\frac{ \CL{e\gamma}{\mu e}^\prime  }{    \CL{e\gamma}{e\mu}^\prime    }\right | 
   \simeq \left |\frac{\tilde\alpha_1}{\tilde\beta_1} \right |
   =\frac{7}{100}\, \frac{\alpha_1}{\beta_1}\simeq 5\times 10^{-3}\,,
   \end{align}
   from  Eq.\,\eqref{Wilson12} and Table \ref{tab:parameters}.
  This prediction is contrast to the prediction
   of the  $\bar \mu_R \sigma_{\mu\nu}e_L$ dominant decay
    in the $U(2)$ flavor model \cite{Tanimoto:2023hse}.
     The angular distribution with respect to the muon polarization can distinguish between $\mu^+ \to e^+_L\gamma$ and $\mu^+ \to e^+_R\gamma$
   \cite{Okada:1999zk}.

On the other hand, the constraint of $|d_e|$ is very tight.
As seen  in Eq.\,\eqref{Wilson12},
\begin{align}
\frac{ {\rm Im}\,[\CL{e\gamma}{e\mu(\mu e)}^\prime]  }
{ {\rm Re}\,[\CL{e\gamma}{\mu\mu}^\prime]    } 
\simeq \left |\frac{\tilde\alpha_1}{\tilde\beta_1} \right |
{\rm Im}\, \delta_{\alpha}
\,.
\label{}
\end{align}
 Since the upper-bound of this ratio is $10^{-8}$ as seen in  Eq.\,\eqref{eq:edm}, we have 
 \begin{align}
 {\rm Im}\, \delta_{\alpha}<10^{-6}\,.
 \label{Im-delta}
 \end{align}
 The constraints of Eqs.\,\eqref{Abs-delta} and \eqref{Im-delta}
 suggest us that
 the electron EDM gives stronger constraint than 
 the  $\mu \to e\gamma$ decay for NP
  if the phase of $\delta_{\alpha}$ is of order one.
 Then, the $\mu \to e\gamma$ decay is much suppressed compared with
 the present experimental upper-bound.
 This numerical prediction is presented in the next subsection.
 However, if  $\delta_{\alpha}$ is real, 
	 the electron EDM does not give constraint for NP,
	 but its magnitude is a prediction through the
	 next-to-leading contribution,
	 which is omitted in Eq.\,\eqref{Wilson12}.  
 The numerical results are also presented in this case in the next subsections.
}


\subsection{ Electron  $(g-2)_{e}$}

The NP of $(g-2)_{\mu}$ and $(g-2)_{e}$ {appears} in the diagonal
{components} of the Wilson coefficient of the dipole operator
at the mass basis.
We have the ratios of the diagonal coefficients from
Eq.\,\eqref{Wilson12} as:
\begin{eqnarray}
\label{CG-ee}
&&\frac{{\rm Re\,} \CL{e\gamma}{ee}^\prime}{{\rm Re\,}\CL{e\gamma}{\mu \mu}^\prime} \simeq \left |\frac{\tilde\alpha_{1(2)}}{\tilde \beta_{1(2)}}\right |
\simeq 4.9\,(4.8)\times 10^{-3} \,,
\end{eqnarray}
where numerical values of Table \ref{tab:parameters} are put
for  $\beta_{1(2)}/\alpha_{1(2)}$.
These predicted ratios are almost in agreement with the charged lepton mass ratio
$m_e/m_\mu=4.84\times 10^{-3}$.

By inputting the experimental value of Eq.\,\eqref{muon-data}, 
the real part of the Wilson coefficient of the muon $\CL{e\gamma}{\mu \mu}^\prime$
has been  {obtained} as seen  in Eq.\,\eqref{Cmumu-exp} \cite{Isidori:2021gqe}.
Now, we can estimate the magnitude of  the electron $(g-2)_e$ anomaly 
by using the  relation in Eq.\,\eqref{CG-ee} as:
\begin{align}
\Delta a_{e} &= \frac{4 m_e}{e}   \frac{v}{\sqrt 2}\,\frac{1}{\Lambda^2} 
\text{Re} \,[\CL{e\gamma}{ee}^\prime] \simeq 5.8\times 10^{-14}\,.
\end{align}
It is easily seen that $\Delta a_{e}$ 
and  $\Delta a_{\mu}$ are proportional to the lepton {masses squared}.
This result is {in} agreement with the naive scaling $\Delta a_\ell \propto m^2_\ell$\cite{Giudice:2012ms}.

In the electron {anomalous} magnetic moment, 
the  experiments \cite{Hanneke:2008tm} give
\begin{align}
a_e^\mathrm{Exp}=1\, 159\, 652\, 180.73(28)\times 10^{-12}\,,
\end{align}
while 
the SM prediction crucially depends on the input value for the fine-structure constant $\alpha$.  Two latest determination
\cite{Parker:2018vye,Rb:2020} based on 
Cesium and Rubidium atomic recoils differ by more than $5\sigma$.
Those observations lead to the difference {from the} SM prediction:
\begin{align}
&&\Delta a_e^{Cs} &= a_e^\mathrm{Exp} - a_e^\mathrm{SM,CS} = \brackets{-8.8 \pm 3.6} \times 10^{-13}~\,, \nonumber\\
&&\Delta a_e^{Rb} &= a_e^\mathrm{Exp} - a_e^\mathrm{SM,Rb} = \brackets{4.8 \pm 3.0} \times 10^{-13}~\,.
\end{align}
Our predicted value is smaller of order one than the  observed one at present. 
We need the precise observation of the fine structure constant
to test our model.

\subsection{Electron EDM and {$\mu\to e\gamma$} decay}

The  LFV  process is severely constrained by 
the experimental  bound 
{$\mathcal{B}\!\brackets{\mu^+ \to e^+ \gamma}<3.1\times 10 ^{-13}$ by the combination of the MEG and MEG II experiments~\cite{TheMEG:2016wtm,MEGII:2023ltw}.}
As seen in subsection \ref{constraint-delta},
the parameter $\delta_{\alpha}$  and
 $\delta_{\beta}$ are constrained in Eq.\,\eqref{Abs-delta}.
On the other hand, the constraint of $|d_e|$ 
by the JILA experiment \cite{Roussy:2022cmp} is much tight
compared with the one from $\mathcal{B}\!\brackets{\mu^+ \to e^+ \gamma}$
as seen  in Eq.\,\eqref{Im-delta}.

At first, suppose the phase of $\delta_{\alpha}$ being  of order one.
Then, the upper-bound of the absolute value of $\delta_{\alpha}$  is
around $ 10^{-6}$ as seen in Eq.\,\eqref{Im-delta}.
Since $\delta_{\beta}$ and $\delta_{\gamma}$ are 
also of the same order $ 10^{-6}$ (See  Eq.\,\eqref{deviation}), we can estimate the branching ratio of the $\mu\to e\gamma$ decay
by taking  $\sigma\simeq|\delta_{\alpha}|\simeq|\delta_{\beta}|\simeq|\delta_{\gamma}|$
as presented in Appendix \ref{distribution}.


\begin{figure}[h!]
	\begin{tabular}{ccc}
		\begin{minipage}{0.47\hsize}
			\begin{center}
				\includegraphics[width=55.0mm]{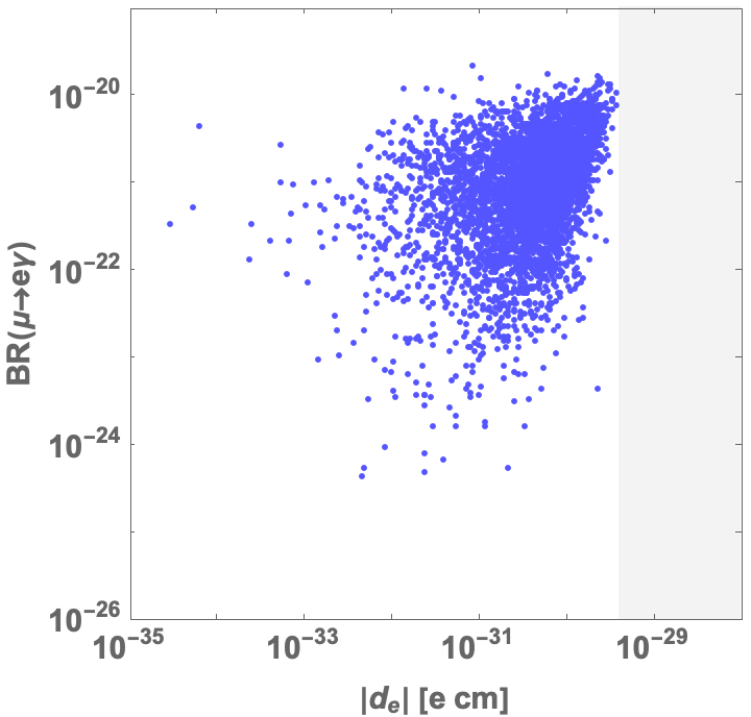}
			  \end{center} \vspace{-0.8cm}
			\caption{Plot of $\mathcal{B}\!\brackets{\mu \to e \gamma}$
				versus $|d_e|$ in Model I,
				where $\sigma=10^{-6}$ is put.
				The grey region is excluded by the experimental upper-bound of $|d_e|$.}
			\label{fig-model1-eEDM-Br}
		\end{minipage}
		\hskip 0.5 cm
		\begin{minipage}{0.47\linewidth}
		\begin{center} 
			\includegraphics[width=55.0mm]{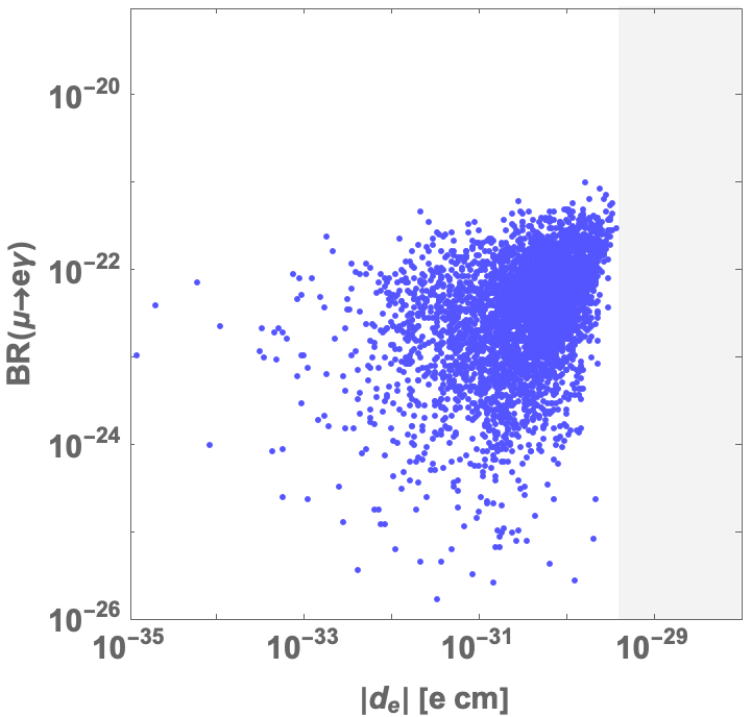}
			 \end{center} \vspace{-0.8cm}
			\caption{Plot of $\mathcal{B}\!\brackets{\mu \to e \gamma}$
				versus $|d_e|$ in Model  {I\hskip-0.4mm I},
				where $\sigma=10^{-6}$ is put.
				The grey region is excluded by the experimental upper-bound of $|d_e|$.}
			\label{fig-model2-eEDM-Br}
		\end{minipage}
	\end{tabular}
\end{figure}
\begin{figure}[h!]
	\begin{tabular}{ccc}
		\begin{minipage}{0.47\hsize}
			\begin{center} 
				\includegraphics[width=55.0mm]{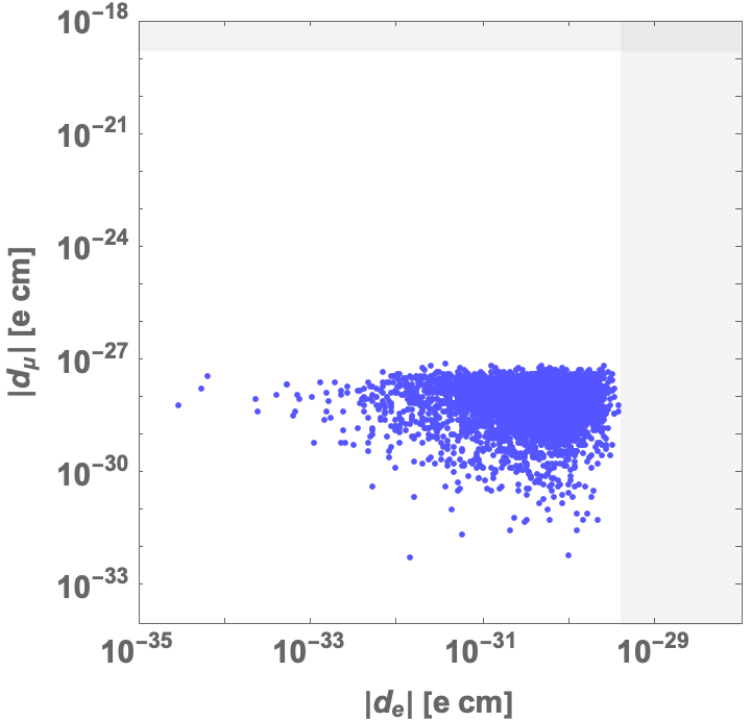} 
		\end{center} \vspace{-0.8cm}
			\caption{Plot of $|d_\mu|$
				versus $|d_e|$ in Model I where $\sigma=10^{-6}$
				is put.
				The grey regions are excluded by the experimental upper-bounds of $|d_e|$ and $|d_\mu|$.}
			\label{fig-model1-e-mu-EDM}
		\end{minipage}
		\hskip 0.5 cm
		\begin{minipage}{0.47\linewidth}
		\begin{center}
				\includegraphics[width=55.0mm]{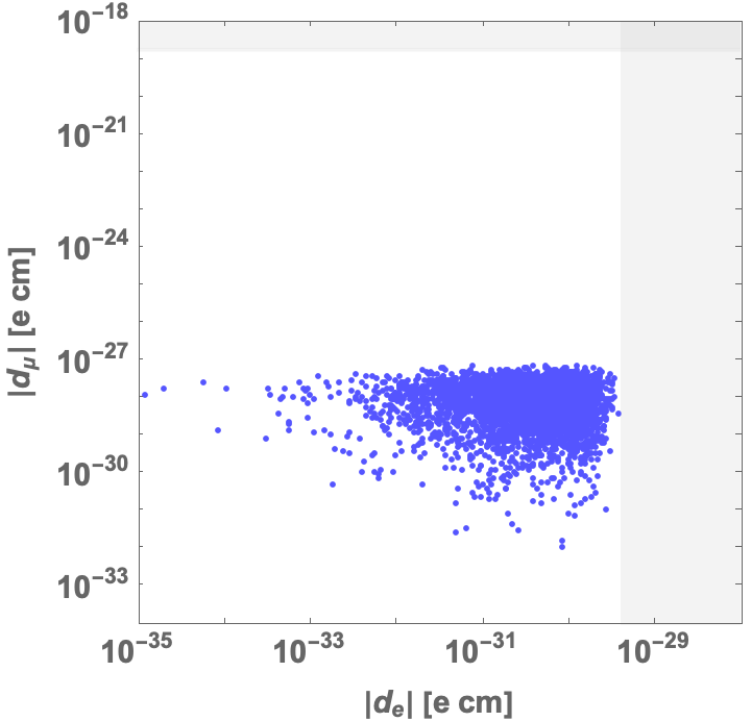} 
	\end{center} \vspace{-0.8cm}
			\caption{Plot of $|d_\mu|$
				versus $|d_e|$ in Model {I\hskip -0.4mm I}, where $\sigma=10^{-6}$ is put.
				The grey regions are excluded by the experimental upper-bounds of $|d_e|$ and $|d_\mu|$.}
			\label{fig-model2-e-mu-EDM}
		\end{minipage}
	\end{tabular}
\end{figure}

In Fig.\,\ref{fig-model1-eEDM-Br},
we plot $\mathcal{B}\!\brackets{\mu \to e \gamma}$
versus the electron EDM $|d_e|$ taking $\sigma= 10^{-6}$ in Model I.
It is found that the electron EDM is almost consistent with
the experimental upper-bound.
Then, the branching ratio of the ${\mu \to e \gamma}$
decay is at most $10^{-20}$.
 In Fig.\,\ref{fig-model2-eEDM-Br},
 we  also plot $\mathcal{B}\!\brackets{\mu \to e \gamma}$
 versus the electron EDM $|d_e|$ taking $\sigma= 10^{-6}$
 in Model  {I\hskip -0.4mm I}.
It is found that
the branching ratio of the ${\mu \to e \gamma}$
 decay is also at most $10^{-21}$.
 There is no hope to observe the ${\mu \to e \gamma}$
 decay in the near future for both models.

 In Fig.\,\ref{fig-model1-e-mu-EDM}, we {show}  the muon EDM $|d_\mu|$ versus the electron EDM $|d_e|$ in Model I.
The predicted upper-bound of  $|d_\mu|$ is around $10^{-27}$e\,cm.
In Fig.\,\ref{fig-model2-e-mu-EDM}, we {also show}  the muon EDM $|d_\mu|$ versus the electron EDM $|d_e|$ in Model {I\hskip -0.4mm I}.
The predicted upper-bound of  $|d_\mu|$ is also  
around $10^{-27}$e\,cm.
In both models,  the ratio of $|d_e/d_\mu|$ is expected to be  the mass ratio $m_e/m_\mu$ approximately.

 	Our prediction of $\mathcal{B}\!\brackets{\mu \to e \gamma}$ depends on the value of $\sigma\simeq |\delta_{\alpha}|
 	\simeq|\delta_\beta|$.
 	The absolute value of $\delta_\alpha$ should be lower than
 	around  $10^{-2}$ from the upper-bound of 
 	$\mathcal{B}\!\brackets{\mu \to e \gamma}$
 	as seen  in Eq.\eqref{Abs-delta}.
 	 While ${\rm Im}\,\delta_{\alpha}$ should be much smaller than  $|\delta_{\alpha}|$ to avoid the constraint of electron EDM.
 	If  $\mathcal{B}\!\brackets{\mu \to e \gamma}$ will be observed
 at the little bit below the present  experimental upper-bound ${\cal O}(10^{-13})$ in the near future, the phase of $\delta_{\alpha}$
 is severely suppressed.
 The simplest expectation is    $\delta_{\alpha}$ being real
 without fine-tuning of the CP phase.
 Leptonic EDMs vanish in both Model I and 
 Model {I\hskip -0.4mm I} under the condition of real  $\delta_{\alpha}$, 
 $\delta_{\beta}$ and  $\delta_{\gamma}$.
 It is remarked that the CP phase of the modular form
 (comes from modulus $\tau$)
 do not contribute to the EDM as far as 
 $\delta_{\alpha}$, 	$\delta_{\beta}$ and  $\delta_{\gamma}$ are real because the $3\times 3$ Wilson coefficient matrix has
 the same phase structure as the  charged lepton mass matrix. 
 
 However,  
 we can consider the alternative case  that $\delta_{\beta}$ and $\delta_{\gamma}$ are complex while
 only $\delta_{\alpha}$ is  real.
 Then,  the imaginary part of $\delta_{\beta}$ can contribute
 to the electron EDM in the next-to-leading order of $\epsilon$.
 Indeed, we find 
 \begin{align}
 |d_e|\sim \frac{\tilde \alpha_{1(2)}}{\tilde \beta_{1(2)}}
 \,\epsilon\,
 {\rm Im}\, \tilde\beta'_{1(2)}
 ={\tilde \alpha_{1(2)}}
 \,\epsilon\,
 {\rm Im}\, \delta_\beta,
 \ \  |d_\mu|\sim {\rm Im}\, \tilde\beta'_{1(2)}
 = \tilde \beta_{1(2)}{\rm Im}\, \delta_\beta,
 \ \  |d_\tau|\sim {\rm Im}\,\gamma'_{1(2)}
 = \gamma_{1(2)} {\rm Im}\, \delta_\gamma,
 \label{real-delta}
 \end{align}
 where ${\rm Im}\, \delta_\beta\sim |\delta_\beta|$
 and ${\rm Im}\, \delta_\gamma\sim |\delta_\gamma|$.
 Under this set up of those  phases,
 we  predict the leptonic EDM in Model I and 
 Model {I\hskip -0.4mm I}.

 \begin{figure}[t!]
 	\begin{tabular}{ccc}
 		\begin{minipage}{0.47\hsize}
 			 \begin{center}
 			\includegraphics[width=55.0mm]{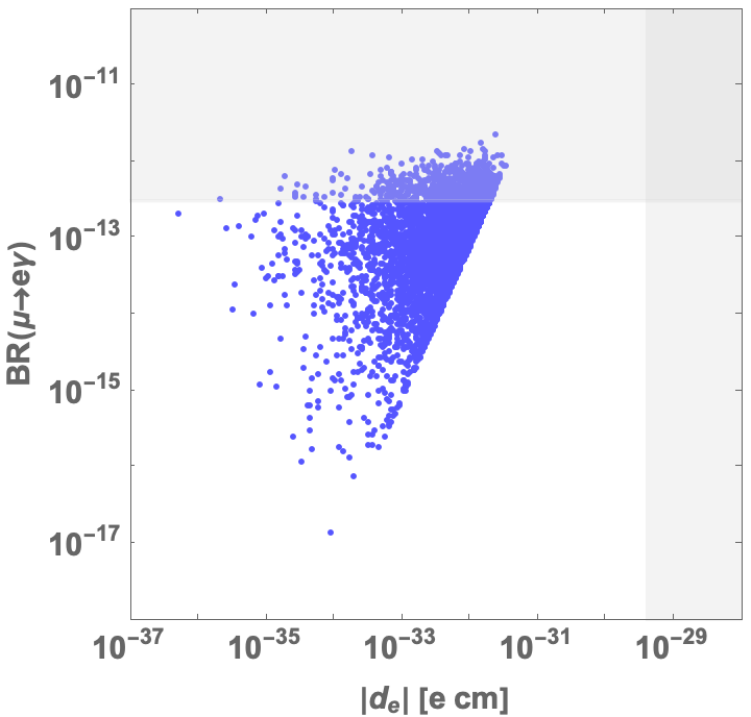} 
 		 \end{center} \vspace{-0.8cm}
 			\caption{Branching ratios of 
 				$\mu\to e\gamma$ versus 	$|d_e|$
 				in Model I with $\sigma=10^{-2}$
 				and ${\rm Im\, \delta_{\alpha}}=0$.
 			The grey regions are excluded by the experimental upper-bounds.}
 			\label{fig:model1-real-e-edm}
 		\end{minipage}
 		\phantom{=}
 		\begin{minipage}{0.47\linewidth}
 			 \begin{center}
 		\includegraphics[width=55.0mm]{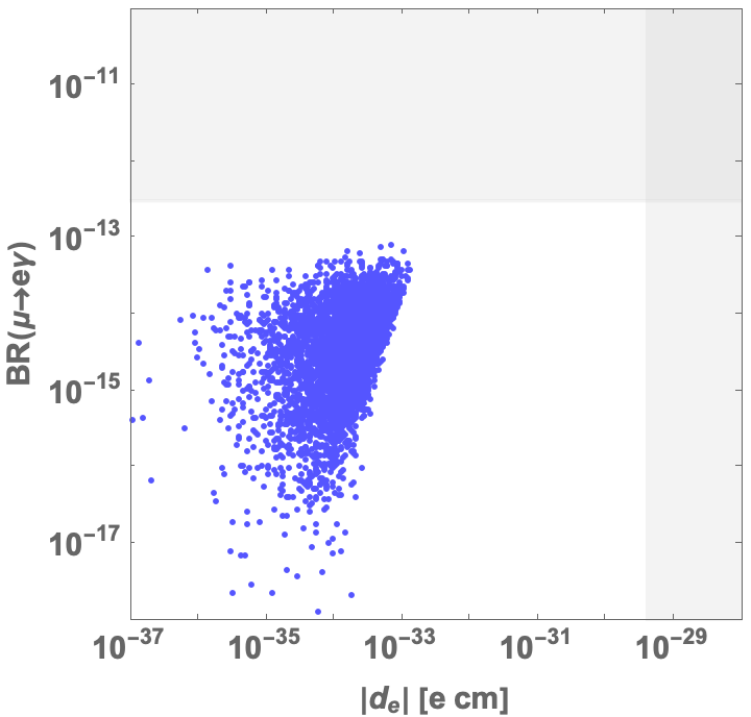} 
 \end{center} \vspace{-0.8cm}
 			\caption{Branching ratios of 
 				$\mu\to e\gamma$ versus 	$|d_e|$
 				in Model {I\hskip-0.4mm I } with $\sigma=10^{-2}$
 				and ${\rm Im\,\delta_{\alpha}}=0$.
 			The grey regions are excluded by the experimental upper-bounds.}
 			\label{fig:model2-real-e-edm}
 		\end{minipage}
 	\end{tabular}
 \end{figure}
 
 In Fig.\,\ref{fig:model1-real-e-edm}, we show
 $\mathcal{B}\!\brackets{\mu \to e \gamma}$
 versus $|d_e|$ taking $\sigma= 10^{-2}$
 with real $\delta_{\alpha}$ in Model I.
 It is found that the electron EDM is almost below $10^{-32}$e\,cm,
 which is of two order smaller than  the present experimental upper bound, while $\mathcal{B}\!\brackets{\mu \to e \gamma}$ is
 close to the experimental present upper-bound.
 In Fig.\,\ref{fig:model2-real-e-edm},
 we  also plot $\mathcal{B}\!\brackets{\mu \to e \gamma}$
 versus  $|d_e|$ taking $\sigma= 10^{-2}$
 with real $\delta_{\alpha}$
 in Model  {I\hskip -0.4mm I}.
 It is found that
 the branching ration of the ${\mu \to e \gamma}$
 and the electron EDM are smaller than of one order than 
 the ones in Model I.

 \begin{figure}[H]
 	\begin{tabular}{ccc}
 		\begin{minipage}{0.47\hsize}
 		\begin{center}
 				\includegraphics[width=55.0mm]{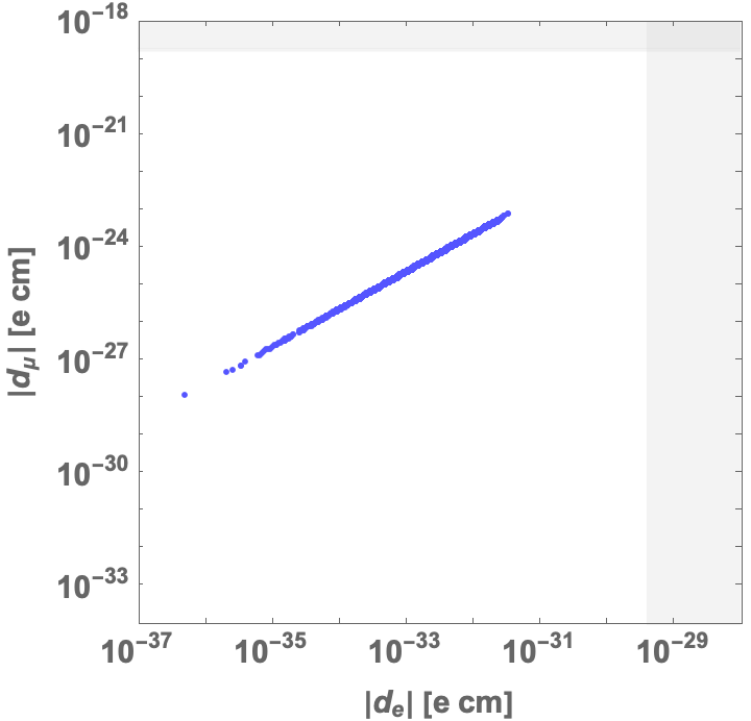} 
 	\end{center} \vspace{-0.8cm}
 			\caption{
 				The muon EDM  versus  electron EDM
 				in Model I with $\sigma=10^{-2}$
 				and ${\rm Im\, \delta_{\alpha}}=0$.
 			The grey regions are  excluded by the experimental upper-bounds.}
 			\label{fig:model1-real-e-mu-edm}
 		\end{minipage}
 		\phantom{=}
 		\begin{minipage}{0.47\linewidth}
 		\begin{center}
 				\includegraphics[width=55.0mm]{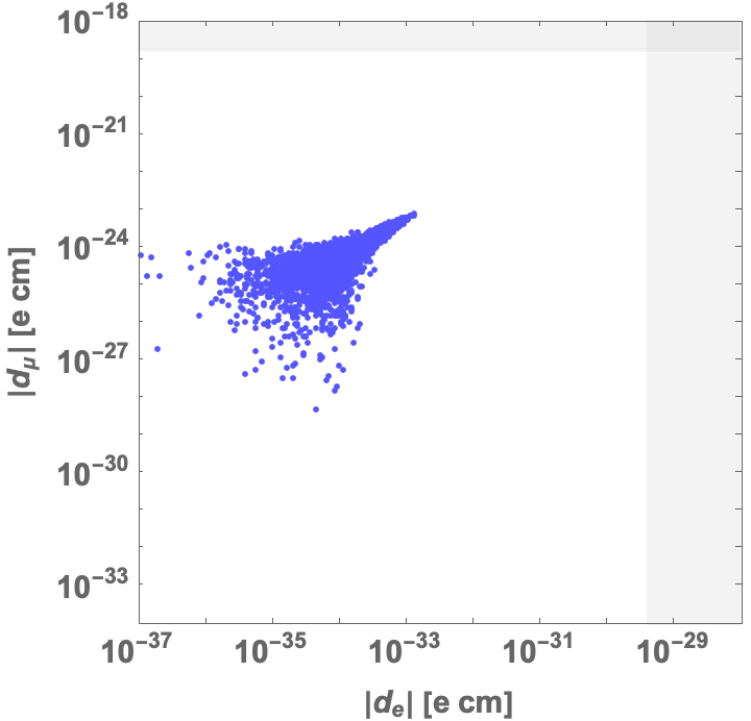} 
 		\end{center} \vspace{-0.8cm}
 			\caption{	The muon EDM  versus  electron EDM
 				in  Model {I\hskip-0.4mm I } with $\sigma=10^{-2}$
 				and ${\rm Im\,\delta_{\alpha}}=0$.
 			The grey regions are excluded by the experimental upper-bounds.}
 			\label{fig:model2-real-e-mu-edm}
 		\end{minipage}
 	\end{tabular}
 \end{figure}
 
 \begin{figure}[H]
 	\begin{tabular}{ccc}
 		\begin{minipage}{0.47\hsize}
 		\begin{center}
 				\includegraphics[width=55.0mm]{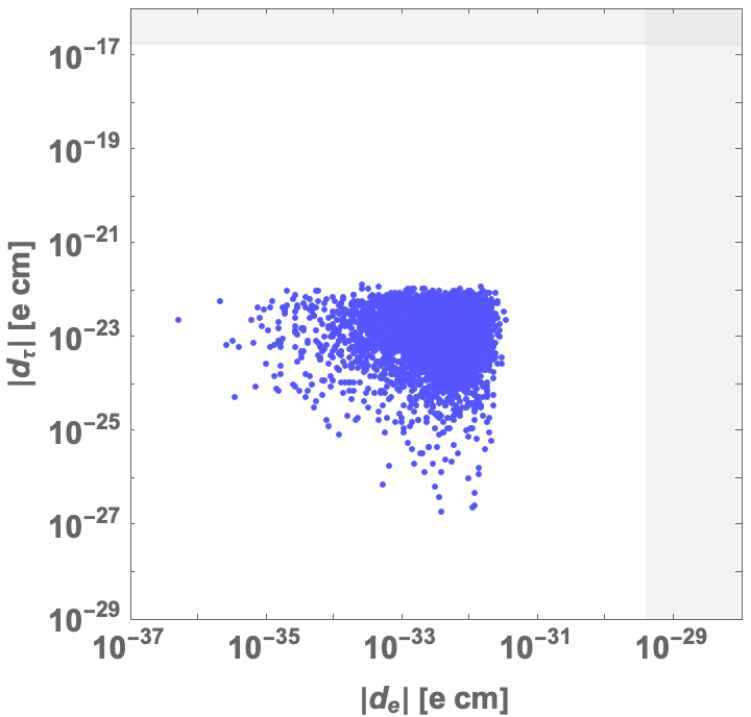}  
 	\end{center} \vspace{-0.8cm}
 			\caption{
 				The tauon EDM  versus  electron EDM
 				in Model I with $\sigma=10^{-2}$
 				and ${\rm Im\, \delta_{\alpha}}=0$.
 			The grey regions are excluded by the experimental upper-bounds.}
 			\label{fig:model1-real-e-tau-edm}
 		\end{minipage}
 		\phantom{=}
 		\begin{minipage}{0.47\linewidth}
 		\begin{center}
 			\includegraphics[width=55.0mm]{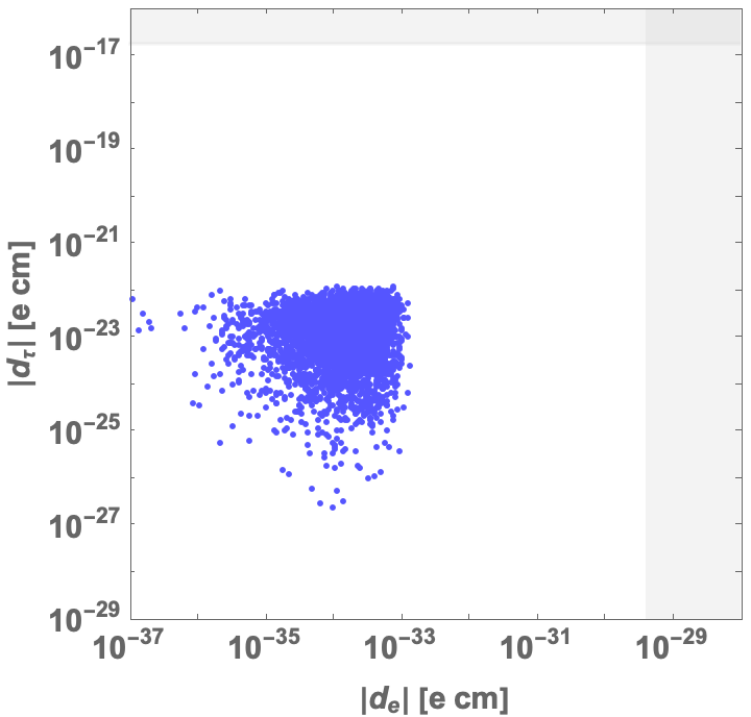} 
 	 \end{center} \vspace{-0.8cm}
 			\caption{	The tauon EDM  versus  electron EDM
 				in  Model {I\hskip-0.4mm I } with $\sigma=10^{-2}$
 				and ${\rm Im\,\delta_{\alpha}}=0$.
 			The grey regions are excluded by the experimental upper-bounds.}
 			\label{fig:model2-real-e-tau-edm}
 		\end{minipage}
 	\end{tabular}
 \end{figure}
 
 In Fig.\,\ref{fig:model1-real-e-mu-edm}, we show also  the muon EDM, $d_\mu$ versus  $|d_e|$ in Model I.
 The predicted upper-bound of  $|d_\mu|$ is
 rather large  around $10^{-23}$e\,cm.
 The both EDMs are almost proportinal to each other,
 which is expected in Eq.\,\eqref{real-delta}.
 In Fig.\,\ref{fig:model2-real-e-mu-edm}, we show   the muon EDM $d_\mu$ versus  $|d_e|$ in Model {I\hskip -0.4mm I}.
 The predicted upper-bound of  $|d_\mu|$ is also  around $10^{-23}$e\,cm.

 The tauon EDM $d_\tau$ is also predicted versus   $|d_e|$ for  Model I and Model {I\hskip -0.4mm I}
 in Figs.\,\ref{fig:model1-real-e-tau-edm}
 and \ref{fig:model2-real-e-tau-edm}, respectively.
 The rather large tauon EDM is predicted  
 to be $|d_\tau|\simeq 10^{-22}$e\,cm
 compared with the muon EDM  because  the ratio of $|d_\mu/d_\tau|$ is expected to be  the mass ratio $m_\mu/m_\tau$ approximately.
 \subsection{LFV decays of tauon in Model {I\hskip-0.4mm I } }

 \begin{figure}[t!]
 	\begin{tabular}{ccc}
 		\begin{minipage}{0.47\hsize}
 	\begin{center}	
 		\includegraphics[width=55.0mm]{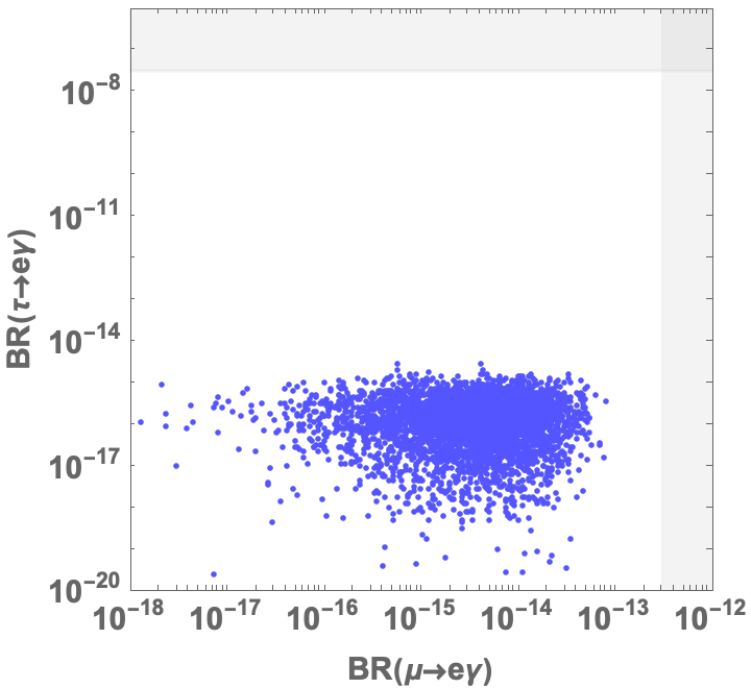}
   \end{center} \vspace{-0.8cm}
 			\caption{Branching ratios of 
 				$\tau\to e\gamma$  versus  	$\mu\to e\gamma$ in Model 
 				I\hskip-0.4mm I with $\sigma=10^{-2}$
 				and ${\rm Im\,\delta_{\alpha}}=0$.
 			The grey regions are excluded by the experimental upper-bounds.}
 			\label{fig:model2-real-tau-e-decay}
 		\end{minipage}
 		\phantom{=}
 		\begin{minipage}{0.47\linewidth}
 	\begin{center}	
 		\includegraphics[width=55.0mm]{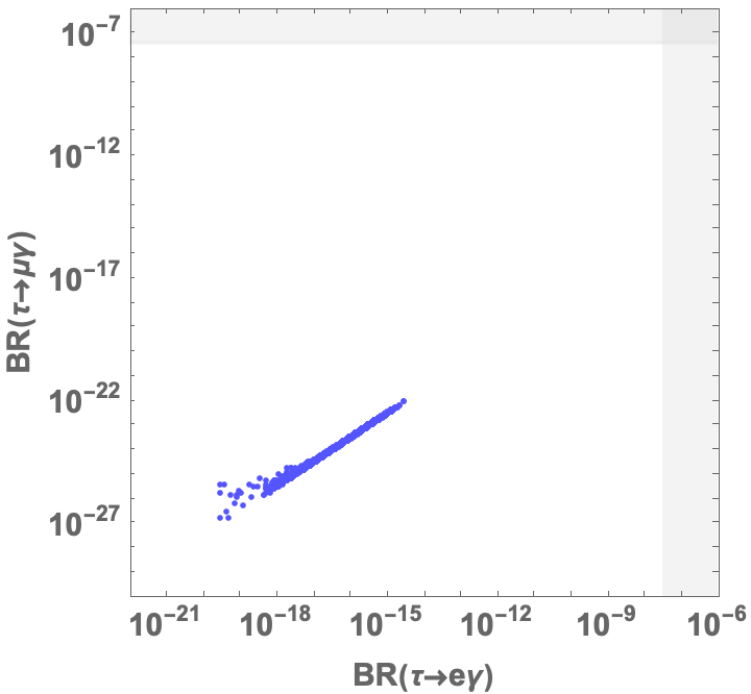} 
  \end{center} \vspace{-0.8cm}
 			\caption{Branching ratios of 
 				$\tau\to e\gamma$ and 	$\tau\to \mu\gamma$ in Model 
 				I\hskip-0.4mm I with $\sigma=10^{-2}$
 				 and ${\rm Im\,\delta_{\alpha}}=0$.
 			 The grey regions are excluded by the experimental upper-bounds.}
 			\label{fig:model2-real-tau-decay}
 		\end{minipage}
 	\end{tabular}
 \end{figure}
 In Model I, $\tau\to e\gamma$ and   $\tau\to \mu\gamma$  decays
 never occur because the tauon decouples to the muon and electron
 as seen in  Eq.\,\eqref{model-I}.
 On the other hand,  in Model {I\hskip-0.4mm I },
 the tauon couples directly the electron as seen in  Eq.\,\eqref{model-II}. Therefore, it also couples to the muon
 in the next-to-leading order.
 Indeed, we can predict the $\tau\to e\gamma$ and  $\tau\to\mu\gamma$  decays numerically.
 Let us   take  $\sigma=10^{-2}$, where
 $\sigma\simeq|\delta_{\alpha}|\simeq |\delta_{\beta}|\simeq |\delta_{\gamma}|$ with ${\rm Im\,\delta_{\alpha}}=0$.
 In Fig.\,\ref{fig:model2-real-tau-e-decay}, we plot 
 the branching ratios of 
 $\tau\to e\gamma$  versus  	$\mu\to e\gamma$ in Model 
 I\hskip-0.4mm I.
 The branching ratios of 
 $\tau\to e\gamma$ is expected  to be
 $10^{-15}$.
 
 In Fig.\,\ref{fig:model2-real-tau-decay},
 we  plot
 the branching ratios of 
 $\tau\to e\gamma$ and 	$\tau\to \mu\gamma$.
 with $\sigma=10^{-2}$
 and ${\rm Im\,\delta_{\alpha}}=0$.
 The branching ratio of $\tau\to \mu\gamma$ is almost proportional to the one of $\tau\to e\gamma$.
 This behavior is understandable that the $\tau-\mu$ coupling
 is induced through $\tau-e$ coupling  in Model {I\hskip-0.4mm I}.
 The branching ratio is at most $10^{-22}$, which is far from the present experimental upper-bound.

\section{Summary}
\label{sec:summary}

We have studied   the leptonic EDM  and the  LFV decays  
relating with the recent data of  anomalous magnetic moment $(g-2)_{\mu}$  in the leptonic dipole operator. 
In order to control the 4-point couplings of SMEFT,
we employ the relation Eq.(\ref{eq:Ansatz}) as {\it Stringy Ansatz},
that is,  higher-dimensional operators are related with 3-point couplings.
We have adopted  the $\Gamma_2$ modular invariant model
to control the flavor structure of leptons, which gives the successful lepton mass matrices \cite{Meloni:2023aru}.
There are two-type mass matrices for the  charged leptons.
The first one is the minimal model, where
 the tauon decouples to the muon and the electron.
 In the  second one,
 the tauon directly couples to the electron.


Suppose the anomaly of the anomalous magnetic moment of the muon
$\Delta a_\mu$ to be  evidence of NP, we have  related it with 
the anomalous magnetic moment of the electron $\Delta a_e$,
 the electron  EDM $d_e$ and the $\mu\to e\gamma$ decay.
It is found that the NP contribution to $a_{e}$
and $a_{\mu}$ {is} proportional to the lepton {masses squared} likewise  the naive scaling 
$\Delta a_\ell \propto m^2_\ell$.
The predicted value of the anomaly of $(g-2)_{e}$ is small of one order compared with the observed one at present. 


It is also remarked that 
 the NP signal of the $\mu\to e\gamma$ process comes from the  operator $\bar e_R \sigma_{\mu\nu}\mu_L$ 
in our scheme.
This prediction is contrast to the one
	in the $U(2)$ flavor model \cite{Tanimoto:2023hse}.

The constraint of $|d_e|$ 
by the JILA experiment \cite{Roussy:2022cmp} is much tight
compared with the one from $\mathcal{B}\!\brackets{\mu^+ \to e^+ \gamma}$
in our framework.
Supposing  the phase of $\delta_{\alpha}$ being  of order one,
 the upper-bound of the absolute value of $\delta_{\alpha}$  is
around $ 10^{-6}$.
Since $\delta_{\beta}$ and $\delta_{\gamma}$ are 
also of the same order $ 10^{-6}$, we can estimate the branching ratio of the $\mu\to e\gamma$ decay.
Then, the branching ration of the ${\mu \to e \gamma}$
decay is at most $10^{-20}$\,($10^{-21}$) 
in Model I ({I\hskip -0.4mm I}).

The smallness of $|d_e|$ comes 
 from the tiny  ${\rm Im}\,\delta_{\alpha}$. 
If  $\delta_{\alpha}$, 
$\delta_{\beta}$ and  $\delta_{\gamma}$ are real, 
leptonic EDMs vanish in both models 
	since the CP phase of the modular form due to  modulus $\tau$
	do not contribute to  EDMs. 
{However, there is a possiblity to {obtain} 
	 $\mathcal{B}\!\brackets{\mu \to e \gamma}\simeq 10^{-13}$
	 while  non-vanishing electron EDM. 
	We have considered  the case  that $\delta_{\beta}$ and $\delta_{\gamma}$ are complex while only $\delta_{\alpha}$ is  real with
 $|\delta_{\alpha}|\simeq |\delta_{\beta}|\simeq|\delta_{\gamma}|\simeq 0.01$.}
Then,  the imaginary part of $\delta_{\beta}$ can contribute
to the electron EDM in the next-to-leading order of $\epsilon$.
The predicted electron EDM is below $10^{-32}$e\,cm,
 while $\mathcal{B}\!\brackets{\mu \to e \gamma}$ is
close to the experimental present upper-bound in Model I.
In Model  {I\hskip -0.4mm I},
the branching ration of the ${\mu \to e \gamma}$
and the magnitude of  the electron EDM are smaller than of one order than  the ones in Model I.
The predicted upper-bound of  $|d_\mu|$ is rather large, around $10^{-23}$e\,cm
and  of  $|d_\tau|$ is $ 10^{-22}$e\,cm in both models.

 Our prediction of the electron EDM is compared with
	other ones of models with the non-Abelian flavor symmetry.
	In  the $\Gamma_3$ ($A_4$) modular invariant model,
	the constraint of $|d_e|$  
	is much looser  than our result.
	Indeed,  $\mathcal{B}\!\brackets{\mu \to e \gamma}$
	is bounded lower than ${\cal O}(10^{-16})$
	 by imposing the experimental constraint of $|d_e|$ \cite{Kobayashi:2022jvy},
	 while it is bounded  lower than ${\cal O}(10^{-20})$ in our model.
	On the other hand, 
	$|d_e|$ is predicted near the present upper-bound, 
	${\cal O}(10^{-30}-10^{-31})$\,e\,cm
	in the  $U(2)_L\otimes U(2)_R$ flavor symmetry \cite{Tanimoto:2023hse}.
	 Thus, the prediction of 	$|d_e|$ depends on the flavor symmetry.
	 In other words, the observation of the electron EDM is good test
	  to distinguish  models with flavor symmetry.

	In Model I, $\tau\to e\gamma$ and   $\tau\to \mu\gamma$  decays
never occur because the tauon decouples to the muon and electron
as seen in  Eq.\,\eqref{model-I}.
On the other hand,  in Model {I\hskip-0.4mm I},
the tauon couples directly the electron as seen in  Eq.\,\eqref{model-II}. Therefore, it also couples to the muon
in the next-to-leading order.
Taking   $\sigma=10^{-2}$ in Model {I\hskip-0.4mm I },
the branching ratios of 
$\tau\to e\gamma$ is expected  to be $10^{-15}$ 
and  of	$\tau\to \mu\gamma$ is at most $10^{-22}$,
which are far from   the present experimental upper-bound.

In our numerical analyses, we do not include the renormalization group (RG) contribution.
The  RG evolution contribution  of the leptonic dipole {operators} 
has been discussed  in Ref.\cite{Isidori:2021gqe,Kobayashi:2022jvy}  to estimate the RG effect  on the numerical results at the low-energy.
Our numerical result
is not so changed even if the RG effect is included
as well as  in Ref.\cite{Kobayashi:2022jvy} .

Finally, it is noted that alternative  lepton flavor model with  the $\Gamma_2$ modular symmetry \cite{Marciano:2024nwm},
which is $''$Seesaw Model$''$ in Table \ref{tab:parameters}
 leads to the similar result to the one of Model I.
Thus,   the modular flavor symmetry 
is powerful to investigate NP of leptons in the framework of SMEFT 
if the symmetry is specified.

\section*{Acknowledgments}
The work was supported by the Fundamental Research Funds for the Central Universities (T.~N.). 

\appendix
\section*{Appendix}
\section{Tensor product of  $\rm S_3$ group}
\label{Tensor}
%

We take the generators of $S_3$ group for the doublet as follows:
\begin{align}
\begin{aligned}
S=\frac{1}{2}
\begin{pmatrix}
-1 & -\sqrt{3} \\
-\sqrt{3} &1 
\end{pmatrix},
\end{aligned}
\qquad \quad
\begin{aligned}
T=
\begin{pmatrix}
1 & 0 \\
0 &-1 
\end{pmatrix}.
\end{aligned}
\label{ST}
\end{align}
%
In this basis, the multiplication rules are:
\begin{align}
& \begin{pmatrix}
a_1\\
a_2
\end{pmatrix}_{\bf 2}
\otimes 
\begin{pmatrix}
b_1\\
b_2
\end{pmatrix}_{\bf 2} 
\nonumber \\
&=\left (a_1 b_1 + a_2 b_2\right )_{\bf 1} 
\oplus \left (a_1 b_2-a_2b_1\right )_{{\bf 1}'}
\oplus \frac13
\begin{pmatrix}
-a_1 b_1 +a_2b_2 \\
a_1 b_2+a_2 b_1 
\end{pmatrix}_{{\bf 2}}
,\\
& {\bf 1} \otimes {\bf 1} = {\bf 1}\,, \qquad 
{\bf 1'} \otimes {\bf 1'} = {\bf 1}\,,
\nonumber \\
& {\bf 1'} \otimes
\begin{pmatrix}
a_1\\
a_2
\end{pmatrix}_{\bf 2}
= 
\begin{pmatrix}
-a_2\\
a_1
\end{pmatrix}_{\bf 2}\,, \
\end{align}
%
where
\begin{align}
S({\bf 1)}=1\,,\qquad S({\bf 1')}=1\,,
\qquad T({\bf 1')}=1\,,\qquad T({\bf 1')}=-1\,.
\label{singlet-charge}
\end{align}
%
Further details can be found in the reviews~
\cite{Ishimori:2010au,Ishimori:2012zz,Kobayashi:2022moq}.
\section{Experimental constraints on the dipole operators}
\label{exp}
From the experimental data of  
the muon $(g-2)_\mu$ and ${\mu \to e\gamma}$,
Ref.\cite{Isidori:2021gqe} gave the  constraints on the dipole operators.
We summarize briefly  them on the dipole operators in Eq.\,\eqref{dipole-operators}.
Below the scale of electroweak symmetry breaking,
the leptonic dipole operators are given as:
\begin{align}
\mathcal{O}_{\underset{rs}{e\gamma}}
= \frac{v }{\sqrt{2}}  \overline{e}_{R_r}  \sigma^{\mu\nu} e_{L_s} F_{\mu\nu}\,,
\label{eq:dipoledef}
\end{align}
where $\{r,s\}$ are  {flavor} indices $e,\mu,\tau$ and $F_{\mu\nu}$ is the electromagnetic field strength tensor.
The effective Lagrangian is
\begin{align}
\mathcal{L}_{\rm dipole}=
\frac{1}{\Lambda^2}\,\left (
\CL{e\gamma}{rs}^\prime\mathcal{O}_{\underset{rs}{e\gamma}}
+\CL{e\gamma}{rs}^\prime\mathcal{O}_{\underset{rs}{e\gamma}}
\right )
\,,
\end{align}
where $\Lambda$ is a certain mass scale of NP  in the effective theory. 
The corresponding Wilson coefficient $\CL{e\gamma}{rs}^\prime$
is denoted in the mass-eigenstate basis of leptons.

The tree-level expression for  $\Delta a_\mu$ in terms of the Wilson coefficient of the dipole operator is 
\begin{align}
\Delta a_{\mu} &= \frac{4 m_{\mu}}{e}   \frac{v}{\sqrt 2} \,\frac{1}{\Lambda^2}\text{Re} \, [\CL{e\gamma}{\mu\mu}^\prime] \,,
\label{eq:magnetic-moment}
\end{align}
where
$v\approx 246$~GeV.
Let us input the value
\begin{align}
\Delta a_\mu = 249 \times 10^{-11}~\,,
\label{appen-muon-input}
\end{align}
then, we obtain the Willson coefficient as:
\begin{align}
\frac{1}{\Lambda^2}\text{Re}\  [\CL{e\gamma}{\mu\mu}^\prime]
= 1.0 \times 10^{-5} \, \mathrm{TeV}^{-2} \,, 
\label{Cmumuappen}
\end{align}
where  $e\simeq 0.3028$ is put in the natural unit.

The tree-level expression of a  radiative LFV rate in terms of the Wilson coefficients  is 
\begin{align}
\mathcal{B}(\ell_r \to \ell_s \gamma) = \frac{m_{\ell_r}^3 v^2}{8 \pi \Gamma_{\ell_r}} \frac{1}{\Lambda^4}\left(|\CL{e\gamma}{rs}'|^2 + |\CL{e\gamma}{sr}'|^2\right) \, .
\label{eq:Branching-ratio_lepton-decay}
\end{align}
Taking the experimental bound { $\mathcal{B}\!\brackets{\mu^+ \to e^+ \gamma} < 3.1 \times 10^{-13}$~(90\%~C.L.) 
obtained by the combination of data from MEG and MEG II experiments~\cite{TheMEG:2016wtm,MEGII:2023ltw} }
in Eq.\,\eqref{LFV-input},
we obtain  the upper-bound of the Wilson coefficient as:
\begin{align}
\frac{1}{\Lambda^2}|\CL{e\gamma}{e\mu(\mu e)}^\prime| < 
  1.8 \times 10^{-10} \, \mathrm{TeV}^{-2} \, .
\label{eq:bound_C_egamma_12-appen}
\end{align}

On the other hand,
by taking the following  experimental upper-bound of the branching ratios, $\mathcal{B}\!\brackets{\tau\to\mu+ \gamma} < 4.2 \times 10^{-8}$ and 
$\mathcal{B}\!\brackets{\tau\to e \gamma} < 3.3 \times 10^{-8}$
\cite{BaBar:2009hkt,Belle:2021ysv},
we obtain  the upper-bound of the Wilson coefficient as:
\begin{align}
\frac{1}{\Lambda^2}|\CL{e\gamma}{\mu\tau(\tau\mu)}^\prime| <  2.65 \times 10^{-6} \, \mathrm{TeV}^{-2} \,, \qquad 
\frac{1}{\Lambda^2}|\CL{e\gamma}{e\tau(\tau e)}^\prime| <  2.35 \times 10^{-6} \, \mathrm{TeV}^{-2}  \,,
\label{eq:bound_C_egamma_23-13-appen}
\end{align}
respectively.

\section{$U_{L1}$ and $U_{R1}$ in Model I}
\label{app-1:UL-UR}
We discuss the charged lepton mass matrix in Model I:
 \begin{align}
{\rm I :} \quad  M_E \simeq v_d
\begin{pmatrix}
A_1 & X_1 p' &0 \\
Y_1 p'&B_1 &0 \\
0&0&\gamma_1
\end{pmatrix}_{RL}\,,
\label{ap:model-I-0}
\end{align}
where
\begin{align}
&A_1=-\tilde \alpha_1 (1-144\epsilon p)\simeq 
-\tilde \alpha_1 \, \qquad B_1=-\tilde \beta_1 (1+24\epsilon p)
\simeq -\tilde \beta_1\,, \nonumber\\
&X_1={\color{blue}}16\sqrt{3}\tilde \alpha_1 \sqrt{\epsilon} \,,\quad\ \ \ \
Y_1= {\color{blue}}8\sqrt{3}\tilde \beta_1 \sqrt{\epsilon}\,.
\end{align}
It is written as:
\begin{align}
M_E \simeq v_d  
\begin{pmatrix}
	-\tilde \alpha_1 & X_1e^{i\pi\tau_R} &0 \\
	Y_1 e^{i \pi\tau_R}&-\tilde \beta_1 &0 \\
	0&0&\gamma_1
\end{pmatrix}, 
\label{ap:model-I-1}
\end{align}
 where $\tau_R\equiv {\rm Re}\,\tau$,
 and ${\cal O}(\epsilon)$ is neglected. 
Then, we have 
\begin{align}
  M_E^\dagger M_E \simeq { v_d^2}
\begin{pmatrix}
\tilde \alpha_1^2+ Y_1^2
& { Z_{L1}}&0 \\
{Z_{L1}^*} & \tilde \beta_1^2+ X_1^2 &0 \\
0&0& \gamma_1^2
\end{pmatrix}\,.
\label{ap:MLL-I-0}
\end{align}
Taking $\tilde \alpha_1 \ll \tilde \beta_1$,
we have 
\begin{align}
Z_{L1}& \simeq  -8\sqrt{3}\tilde \beta_1^2\sqrt{\epsilon}
\left (1+2\frac{\tilde \alpha^2_1}{\tilde \beta^2_1}
e^{2i\pi\tau_R}\right )e^{-i\pi\tau_R}\nonumber\\
&\simeq -8\sqrt{3}
\tilde \beta_1^2\sqrt{\epsilon}\left ( 1+2\frac{\tilde \alpha^2_1}{\tilde \beta^2_1}
\cos{2\pi\tau_R}\right )e^{-i\pi\tau_R}
e^{i\phi_{L1}}\,,
\end{align}
where
\begin{align}
\phi_{L1}\simeq 2\frac{\tilde \alpha^2_1}{\tilde \beta^2_1}
\sin{2\pi\tau_R}\,.
\end{align}

By using phase matrix $P_{L1}$, it is rewritten as:
\begin{align}
M_E^\dagger M_E \simeq { v_d^2}
P_{L1}
\begin{pmatrix}
\tilde \alpha_1^2
& -| Z_{L1}| &0 \\
-| Z_{L1}|& \tilde \beta_1^2 &0 \\
0&0& \gamma_1^2
\end{pmatrix} P_{L1}^*\,,
\label{ap:MLL-I-2}
\end{align}
\begin{align}
 P_{L1}=\begin{pmatrix}
	1 & 0 &0 \\
	0&e^{i(\pi\tau_R-\phi_{L1})}&0 \\  0&0&1
\end{pmatrix},
\label{PL1}
\end{align}
where
\begin{align}
|Z_{L1}| = 8\sqrt{3}
\tilde \beta_1^2\sqrt{\epsilon}\left ( 1+2\frac{\tilde \alpha^2_1}{\tilde \beta^2_1}\cos{2\pi\tau_R}\right )\,.
\end{align}
The diagonal form of  $M_E^\dagger M_E$ 
is obtained  by using the following orthogonal matrix $U_{L1}$ 
as
$U_{L1}^T (P_{L1}^*M_E^\dagger M_E P_{L1}) U_{L1}$:
\begin{align}
U_{L1}=\begin{pmatrix}
\cos \theta_{L1} & \sin \theta_{L1} &0 \\
-\sin \theta_{L1}&\cos \theta_{L1}&0 \\  0&0&1
\end{pmatrix}\simeq \begin{pmatrix}
1 &  \theta_{L1} &0 \\ -\theta_{L1}&1&0 \\  0&0&1
\end{pmatrix},
\label{UL1}
\end{align}
where 
\begin{align}
&\tan 2\theta_{L1}=\frac{-2 |Z_{L1}|}
{\tilde \beta_1^2-\tilde \alpha_1^2}
\sim -\frac{2 |Z_{L1}|}
{\tilde \beta_1^2} \left ( 1-\frac{\tilde \alpha^2_1}{\tilde \beta^2_1}\right )^{-1}
 \simeq -16 \sqrt{3}\sqrt{\epsilon}
 \left [ 1+\frac{\tilde \alpha^2_1}{\tilde \beta^2_1}
 \left (1+2\cos{2\pi\tau_R}\right )\right ].
\end{align}
Then, we have 
\begin{align}
\theta_{L1}
\simeq -8 \sqrt{3}\sqrt{\epsilon}
\left [ 1+\frac{\tilde \alpha^2_1}{\tilde \beta^2_1}
\left (1+2\cos{2\pi\tau_R}\right )\right ]\,.
\label{mixingL-1}
\end{align}

 Let us discuss $M_E M_E^\dagger$  to obtain the right-handed mixing:
\begin{align}
M_E M_E^\dagger \simeq { v_d^2} 
\begin{pmatrix}
\tilde \alpha_1^2+ X_1^2
& { Z_{R1}}&0 \\
{Z_{R1}^*} & \tilde \beta_1^2+ Y_1^2 &0 \\
0&0& \gamma_1^2
\end{pmatrix}_{RR} \,,
\label{ap:MRR-I-0}
\end{align}
where
\begin{align}
&Z_{R1} =-\tilde \alpha_1 Y_1 e^{-i\pi\tau_R}
-\tilde \beta_1 X_1 e^{i\pi\tau_R}
=-8\sqrt{3}\tilde\alpha_1 \tilde\beta_1 \sqrt{\epsilon}
(e^{-i\pi\tau_R}+2e^{i\pi\tau_R})
\nonumber\\
&=-8\sqrt{3}\tilde\alpha_1 \tilde\beta_1 \sqrt{\epsilon}
(3\cos\pi\tau_R+i\sin \pi\tau_R)
 =-8\sqrt{3}\tilde\alpha_1 \tilde\beta_1 \sqrt{\epsilon}
\sqrt{5+4\cos 2\pi\tau_R} e^{i\phi_{R}}\,,
\end{align}
where
\begin{align}
\tan \phi_{R}= \frac13\tan\pi\tau_R\,.
\end{align}

By using the phase matrix $P_R$, we have 
\begin{align}
M_E M_E^\dagger \simeq { v_d^2} P_{R1}
\begin{pmatrix}
\tilde \alpha_1^2+ X_1^2
& -| Z_{1R}|&0 \\
-|Z_{1R}|& \tilde \beta_1^2+ Y_1^2 &0 \\
0&0& \gamma_1^2
\end{pmatrix}_{RR} \hskip -0.3 cm P_{R1}^*\,,
\label{ap:MRR-I-1}
\end{align}

\begin{align}
P_{R1}=\begin{pmatrix}
1 & 0 &0 \\
0&e^{-i\phi_{R}}&0 \\  0&0&1
\end{pmatrix}\,.
\label{PR1}
\end{align}

The diagonal form of  $M_E M_E^\dagger $ 
is obtained  by using the following orthogonal matrix $U_R$ as
$U_{R1}^T (P_{R1}^* M_E M_E^\dagger P_{R1}) U_{R1} $:

\begin{align}
U_{R1}=\begin{pmatrix}
\cos \theta_{R1} & \sin \theta_{R1} &0 \\
-\sin \theta_{R1}&\cos \theta_{R1}&0 \\  0&0&1
\end{pmatrix}\simeq \begin{pmatrix}
1 &  \theta_{R1} &0 \\ -\theta_{R1}&1&0 \\  0&0&1
\end{pmatrix},
\label{UR1}
\end{align}
where 
\begin{align}
&\tan 2\theta_{R1}=\frac{-2 |Z_{R1}|}
{\tilde \beta_1^2 -\tilde \alpha_1^2}
\simeq -\frac{2 |Z_{R1}|}
{\tilde \beta_1^2} \simeq -16 \sqrt{3}\sqrt{\epsilon}
\frac{\tilde \alpha_1}{\tilde \beta_1}\sqrt{5+4\cos 2\pi\tau_R}\, .
\end{align}
Finally, we get
 \begin{align}
 \theta_{R1}
 \simeq -8 \sqrt{3}\sqrt{\epsilon}
 \frac{\tilde \alpha_1}{\tilde \beta_1}\sqrt{5+4\cos 2\pi\tau_R}\,.
 \label{mixingR-1}
 \end{align}

In order to obtain the approximate  diagonal  mass matrix  from Eq.\,\eqref{ap:model-I-1}
by using the mixing angles of 
Eqs.\,\eqref{mixingL-1}, \eqref{mixingR-1} and phase matrix
 $P_{L1}$ of Eq.\,\eqref{PL1}, $P_{R1}$ of Eq.\,\eqref{PR1}, 
  we calculate   $U_{R1}^T P_{R1}^* M_E P_{L1} U_{L1}$.
Then we obtain the diagonal matrix elements up to
${\cal O}(\sqrt{\epsilon})$ as follows:

\begin{align}
&M_E(1,1) \simeq  
-\tilde \alpha_1\,, \qquad M_E(2,2) \simeq  
-\tilde \beta_1\,,\qquad M_E(3,3)=\gamma_1\,,
\nonumber\\
&M_E(1,2) \simeq  
8\sqrt{3}\tilde \alpha_1 \sqrt{\epsilon}e^{-i(\phi_{R}+\pi\tau_R)}
\left [1+2e^{2i\pi\tau_R}
-e^{i(\phi_{R}+\pi\tau_R)}\sqrt{5+4\cos 2\pi\tau_R} \right ] = 0\,,
\nonumber\\
&M_E(2,1) \simeq  8\sqrt{3}\tilde \alpha_1 \frac{\tilde \alpha_1}{\tilde \beta_1}\sqrt{\epsilon}
\left[-(1+2 e^{-2i\pi\tau_R})e^{i(\phi_{R}+\pi\tau_R)}+\sqrt{5+4\cos 2\pi\tau_R}\right ]=0\,,
\nonumber\\
&M_E(1,3) =M_E(3,1)=M_E(2,3) =M_E(3,2)=0\, ,
\end{align}
{where the phases of diagonal matrix elements are removed
by the redefinition of charged lepton fields.}
In  $M_E(1,2)$ and $M_E(2,1)$, we have used following equations: 
{
	\begin{align}
 \tan \phi_{R}= \frac13\tan\pi\tau_R\,,
 \qquad 1+2 e^{2i\pi\tau_R}=\sqrt{5+4\cos 2\pi\tau_R}\  e^{i(\phi_{R}+\pi\tau_R)}\,.
\end{align}}

\section{$U_{L2}$ and $U_{R2}$ in Model {\rm {I\hskip -0.4mm I} }}
\label{app-2:UL-UR}
We discuss the charged leptom mass matrix in Model
 {\rm {I\hskip -0.4mm I} }:
\begin{align}
{\rm {I\hskip -0.4mm I} :} \quad  M_E \simeq v_d
\begin{pmatrix}
A_2 & X_2 p' &F_2 p' \\
Y_2 p'&B_2 &0 \\
0&0&\gamma_2
\end{pmatrix}_{RL}\,,
\label{ap:model-II-0}
\end{align}
where

\begin{align}
&A_2=\tilde \alpha_2 (1+264\epsilon p)\simeq 
\tilde \alpha_2 \,,\qquad
B_2=-\tilde \beta_2 (1+24\epsilon p)
\simeq -\tilde \beta_2\,, \nonumber\\
&X_2={\color{blue}}8\sqrt{3}\tilde \alpha_2 \sqrt{\epsilon} \,,\quad
Y_2= {\color{blue}}8\sqrt{3}\tilde \beta_2 \sqrt{\epsilon}\,,\quad
F_2=-24\sqrt{3}\tilde \alpha_D \sqrt{\epsilon}\,.
\end{align}
It is written as: 
\begin{align}
M_E \simeq v_d  
\begin{pmatrix}
\tilde \alpha_2 & X_2 e^{i\pi\tau_R}&F_2 e^{i\pi\tau_R} \\
Y_2 e^{i\pi\tau_R}&-\tilde \beta_2 &0 \\
0&0&\gamma_2
\end{pmatrix}\,, 
\label{ap:model-II-1}
\end{align}
where $\tau_R\equiv {\rm Re}\,\tau$. 
Then,  we have 
\begin{align}
M_E^\dagger M_E \simeq { v_d^2}
\begin{pmatrix}
\tilde \alpha_2^2+ Y_2^2
& { Z_{L2}}&\tilde \alpha_2 F_2\, e^{i\pi\tau_R} \\
{Z_{L2}^*} & \tilde \beta_2^2+ X_2^2 &  X_2 F_2 \\
\tilde \alpha_2 F_2\, e^{-i\pi\tau_R}&X_2 F_2& \gamma_2^2
\end{pmatrix}\,,
\label{ap:MLL-II-0}
\end{align}
where
\begin{align}
&Z_{L2} = -8\sqrt{3}\tilde \beta_2^2\sqrt{\epsilon}e^{-i\pi\tau_R}
\left(1-\frac{\tilde\alpha_2^2}{\tilde\beta_2^2}e^{2i\pi\tau_R}\right )
 \simeq  -8\sqrt{3}\tilde \beta_2^2\sqrt{\epsilon}e^{-i\pi\tau_{R}}
 e^{i\phi_{L2}}
 \left |1-\frac{\tilde\alpha_2^2}{\tilde\beta_2^2}e^{2i\pi\tau_R}\right |\,\nonumber\\
& \simeq 
 -8\sqrt{3}\tilde \beta_2^2\sqrt{\epsilon}e^{-i\pi\tau_R}e^{i\phi_{L2}}
 \left (1- \frac{\tilde\alpha_2^2}{\tilde\beta_2^2}\cos{2\pi\tau_R}\right )
\,,\end{align}
and
\begin{align}
\phi_{L2}\simeq -\frac{\tilde\alpha_2^2}{\tilde\beta_2^2}\sin 2\pi \tau_R\,.
\end{align}

Up to ${\cal O}(\sqrt{\epsilon})$, we have 
\begin{align}
M_E^\dagger M_E \simeq { v_d^2}
\begin{pmatrix}
\tilde \alpha_2^2
& -| Z_{L2}| e^{-i(\pi\tau_R-\phi_{L2})}&\tilde \alpha_2 F_2\, e^{i\pi\tau_R} \\
-| Z_{L2}| e^{i(\pi\tau_R-\phi_{L2})}& \tilde \beta_2^2 &X_2 F_2 \\
\tilde \alpha_2 F_2\, e^{-i\pi\tau_R}&X_2 F_2& \gamma_2^2
\end{pmatrix}\,,
\label{ap:MLL-II-1}
\end{align}
where the (2,3) and (3,2) entries are ${\cal O}({\epsilon})$.
By using phase matrix $P_{L2}$, it is rewritten as:
\begin{align}
M_E^\dagger M_E \simeq { v_d^2}\,
P_{2L}
\begin{pmatrix}
\tilde \alpha_1^2
& -| Z_{L2}| &{-24\sqrt{3}\tilde \alpha_2 \tilde \alpha_D \sqrt{\epsilon}} \\
-| Z_{L2}|& \tilde \beta_2^2 
&-576\,\tilde \alpha_2 \tilde \alpha_D \,\epsilon
 \, {e^{-i(2\pi\tau_R-\phi_{L2})}}\\
{-24\sqrt{3}\tilde \alpha_2 \tilde \alpha_D \sqrt{\epsilon}}&-576\,\tilde \alpha_2 \tilde \alpha_D \,\epsilon
 \, {e^{i(2\pi\tau_R-\phi_{L2})}}& \gamma_2^2
\end{pmatrix} P_{2L}^*,
\label{ap:MLL-II-2}
\end{align}
where
\begin{align}
P_{L2}=\begin{pmatrix}
1&0& 0  \\
0&e^{i(\pi\tau_R-\phi_{L2})} &0 \\  0&0&{e^{-i\pi\tau_R}}
\end{pmatrix}\,,
\label{PL2}
\end{align}
and 
\begin{align}
 |Z_{L2}|\simeq 
8\sqrt{3}\tilde \beta_2^2\sqrt{\epsilon}
\left (1- \frac{\tilde\alpha_2^2}{\tilde\beta_2^2}\cos{2\pi\tau_R}\right )
\,.
\end{align}


The diagonal form of  $M_E^\dagger M_E$ 
is obtained  by using the following unitary matrix $U_{L2}$ as
$U_{L2}^\dagger (P_{L2}^*M_E^\dagger M_E P_{L2})U_{L2} $.
Since (1,3), (3,1), (2,3) and (3,2) elements of   $M_E^\dagger M_E$ 
is much smaller than $\tilde \beta_2^2$ and $|Z_{L2}|$,
it is given approximately:
\begin{align}
U_{L2}\simeq \begin{pmatrix}
\cos \theta_{L2} & \sin \theta_{L2} &\theta_{L13} \\
-\sin \theta_{L2}&\cos \theta_{L2}& V_{L23} \\   -\theta_{L13}& -V^*_{L23}&1
\end{pmatrix}
\simeq \begin{pmatrix}
1 &  \theta_{L2} & \theta_{L13} \\
-\theta_{L2}&1&V_{L23} \\   -\theta_{L13}& -V^*_{L23}&1
\end{pmatrix},
\label{UL2}
\end{align}
where 
\begin{align}
&\tan 2\theta_{L2}=\frac{-2 |Z_{L2}|}
{\tilde \beta_2^2-\tilde \alpha_2^2}
\simeq -\frac{2 |Z_{L2}|}
{\tilde \beta_2^2} \left ( 1-\frac{\tilde \alpha^2_2}{\tilde \beta^2_2}\right )^{-1}
\simeq -16 \sqrt{3}\sqrt{\epsilon}
\left [ 1+\frac{\tilde \alpha^2_2}{\tilde \beta^2_2}
\left (1-\cos{2\pi\tau_R}\right )\right ].
\end{align}
Then, we have 
\begin{align}
\theta_{L2}
\simeq -8 \sqrt{3}\sqrt{\epsilon}
\left [ 1+\frac{\tilde \alpha^2_2}{\tilde \beta^2_2}
\left (1-\cos{2\pi\tau_R}\right )\right ]
\,.
\label{mixingL-2}
\end{align}
On the other hand, we have 
\begin{align}
\theta_{L13}\simeq 
-\frac{24 \sqrt{3}\tilde \alpha_2 \tilde \alpha_D}{\gamma_2^2}
 \, \sqrt{\epsilon}\,,\qquad 
 V_{L23}\simeq 
\frac{576 \tilde \alpha_2 \tilde \alpha_D}{\gamma_2^2}
 \,\epsilon\,\left [1-e^{i(\phi_{L2}-2\pi\tau_R)}\right ].
\end{align}

Let us discuss $M_E M_E^\dagger$ to obtain the right-handed mixing:
\begin{align}
M_E M_E^\dagger \simeq { v_d^2} 
\begin{pmatrix}
\tilde \alpha_2^2+ X_2^2+F_2^2
& { Z_{R2}}& F_2 \gamma_2 p'\\
{Z_{R1}^*} & \tilde \beta_2^2+ Y_2^2 &0 \\
F_2 \gamma_2 p'^*&0& \gamma_2^2
\end{pmatrix}_{RR}\,,
\label{ap:MRR-II-0}
\end{align}
where
\begin{align}
&Z_{R2} =\tilde \alpha_2 Y_2 e^{-i\pi\tau_R}
-\tilde \beta_2 X_2 e^{i\pi\tau_R}
=8\sqrt{3}\tilde\alpha_2 \tilde\beta_2 \sqrt{\epsilon}
(e^{-i\pi\tau_R}-e^{i\pi\tau_R})
=-16i\sqrt{3}\tilde\alpha_2 \tilde\beta_2 \sqrt{\epsilon}
\sin \pi\tau_R\,.
\end{align}

By using the phase matrix $P_{2R}$, we have 
\begin{align}
M_E M_E^\dagger \simeq { v_d^2} P_{R2}
\begin{pmatrix}
\tilde \alpha_2^2(1+192\epsilon)
& -| Z_{R2}|&{-24\sqrt{3}\tilde \alpha_D\gamma_2 \sqrt{\epsilon}} \\
-|Z_{R2}|& \tilde \beta_2^2(1+192\epsilon)& 0 \\
{-24\sqrt{3}\tilde \alpha_D\gamma_2 \sqrt{\epsilon}}&0& \gamma_2^2
\end{pmatrix}_{RR} \hskip -0.3 cm P_{R2}^*\,,
\label{ap:MRR-II-1}
\end{align} where
\begin{align}
P_{R2}=\begin{pmatrix}
1 & 0 &0 \\
0&-i&0 \\  0&0& {e^{-i\pi\tau_R}}
\end{pmatrix},
\label{PR2}
\end{align}
and
\begin{align}
&|Z_{R2}| =16\sqrt{3}\tilde\alpha_2 \tilde\beta_2 \sqrt{\epsilon}
\sin \pi\tau_R\,.
\end{align}

The diagonal form of  $M_E M_E^\dagger $ 
is obtained  by using the following orthogonal matrix $U_{R2}$ as
$U_{R2}^T (P_{R2}^* M_E M_E^\dagger P_{R2}) U_{R2} $:

\begin{align}
U_{R2}\simeq\begin{pmatrix}
\cos \theta_{R2} & \sin \theta_{R2} &\theta_{R13} \\
-\sin \theta_{R2}&\cos \theta_{R2}&0 \\  -\theta_{R13}&0&1
\end{pmatrix}\simeq \begin{pmatrix}
1 &  \theta_{R2} &\theta_{R13} \\ -\theta_{R2}&1&0\\  -\theta_{R13}&0&1
\end{pmatrix},
\label{UR2}
\end{align}
where 
\begin{align}
&\tan 2\theta_{R2}=\frac{-2 |Z_{R2}|}
{\tilde \beta_2^2 -\tilde \alpha_2^2}
\simeq -\frac{2 |Z_{R2}|}
{\tilde \beta_2^2}\simeq -32 \sqrt{3}\sqrt{\epsilon}
\frac{\tilde \alpha_2}{\tilde \beta_2}\sin \pi\tau_R\,.
\end{align}
Therefore, we get approximately
\begin{align}
\theta_{R2}
\simeq -16 \sqrt{3}\sqrt{\epsilon}
\frac{\tilde \alpha_2}{\tilde \beta_2}\sin\pi\tau_R\,.
\label{mixingR-2}
\end{align}
On the other hand,
 we have 
 \begin{align}
 {\theta_{R13}
 \simeq -24\sqrt{3}\frac{\tilde \alpha_D}{\gamma_2} \sqrt{\epsilon}}\,.
 \end{align}

In order to obtain the approximate  diagonal  mass matrix  from Eq.\,\eqref{ap:model-II-1}
by using the mixing angles of 
Eqs.\,\eqref{mixingL-2}, \eqref{mixingR-2} and phase matrix
$P_{L2}$ of Eq.\,\eqref{PL2}, $P_{R2}$ of Eq.\,\eqref{PR2}, 
we calculate   $U_{R2}^T P_{R2}^* M_E P_{L2} U_{L2}$.
Then we obtain the approximate mass matrix elements 
up to ${\cal O}(\sqrt{\epsilon})$ as follows:

\begin{align}
&M_E(1,1) \simeq  
\tilde \alpha_2\,, \qquad M_E(2,2) \simeq  
-\tilde \beta_2\,, \qquad  M_E(3,3)\simeq \gamma_2\,,
\nonumber\\
&M_E(1,2) \simeq  
8\sqrt{3}\tilde \alpha_2 \sqrt{\epsilon}\,
i e^{-i\pi\tau_R}
\left (-1+e^{2i\pi\tau_R}-2i\,e^{i\pi\tau_R}\sin\pi\tau_R\right ) = 0\,,
\nonumber\\
&M_E(2,1) \simeq 8\, i\,e^{i\pi\tau_R}\sqrt{3}\tilde \alpha_2 \frac{\tilde \alpha_2}{\tilde \beta_2}\sqrt{\epsilon}
(e^{-2i\pi\tau_R}-1+2i\sin\pi\tau_R e^{-i\pi\tau_R})=0 \,,
\nonumber\\
&M_E(1,3) \simeq -24\sqrt{3}\tilde \alpha_2 
\frac{\tilde \alpha_2\tilde\alpha_D}{\gamma_2^2}\sqrt{\epsilon}
\,,
\nonumber\\
&M_E(3,1) \simeq - 4608\sqrt{3}\tilde \alpha_2 
\frac{\tilde \alpha_D}{\gamma_2}\epsilon\sqrt{\epsilon}
e^{2i\pi\tau_R}
\,,
\nonumber\\
&M_E(2,3) \sim \tilde \alpha_2\, {\cal O} \left(\frac{\tilde \alpha_D}{ \tilde\beta_2}\epsilon \right) \,,
\nonumber\\
&M_E(3,2) \sim \tilde \alpha_2\, {\cal O} \left(\frac{\tilde \alpha_D}{ \gamma_2}\frac{\tilde \alpha^2_2}{ \tilde\beta^2_2}\epsilon \right) \,,
\end{align}
{where the phases of diagonal matrix elements are removed
	by the redefinition of charged lepton fields.}
In $M_E(1,2)$ and $M_E(2,1)$, we have used the identity:
{
	\begin{align}
	1-e^{2i\pi\tau_R}=-2i\,e^{i\pi\tau_R}\sin\pi\tau_R\,.
	\end{align}}

{
\section{Model parameters in Normal distribution}
\label{distribution}
In our analyses, we scatter the parameters 
$\delta_\alpha$, $\delta_\beta$, $\delta_\gamma$
and $\delta_D$ in the normal distribution with an average $0$
and the standard deviation $\sigma$
in Eq.\,\eqref{deviation}
\begin{eqnarray}
F=F_0 \exp{\left (-\frac{x^2}{2\sigma^2}\right )}\,,
\end{eqnarray}	
where $F_0$ is a normalization constant.
Since the mean square of $x$ is ${<x^2>}=\sigma^2$,
 we have 
 \begin{eqnarray}
\sqrt{ <\delta^2_\alpha>}=\sqrt{<\delta^2_\beta>}
 =\sqrt{<\delta^2_\gamma>}=
 \sqrt{<\delta^2_D>}= \sigma\,.
 \end{eqnarray}	
 In our   statistical discussions, we take
  \begin{eqnarray}
|\delta_\alpha|\simeq  \sqrt{ <\delta^2_\alpha>}\,,
\quad |\delta_\beta|\simeq\sqrt{<\delta^2_\beta>}\,,
\quad |\delta_\gamma|
 \simeq\sqrt{<\delta^2_\gamma>}\,,\quad
 |\delta_D|\simeq\sqrt{<\delta^2_D>}\,.
 \end{eqnarray}	
 Therefore, we obtain  the relevant factors as:
 \begin{eqnarray}
 \left | 1-\frac{\alpha_{1(2)}}{\alpha'_{1(2)}}
 \frac{\beta'_{1(2)}} {\beta_{1(2)}}\right |\simeq \sigma\,,
 \quad \left | 1-\frac{\alpha_{D}}{\alpha'_{D}}
 \frac{\gamma'_{2}} {\gamma_{2}} \right |\simeq \sigma\,,
 \label{sigma}
 \end{eqnarray}	
 which appear in the coefficients of 
 Eqs.\,\eqref{Wilson12} and \eqref{Wilson22}.}


\end{document}